\documentclass[twocolumn]{article}
\makeatletter
\let\ifaastex\@secondoftwo
\let\ifarxiv\@firstoftwo
\makeatother
\ifarxiv{\usepackage[utf8]{inputenc}}{}
\usepackage{font-termes}
\ifaastex{
  \usepackage[hidelinks]{hyperref}
  \usepackage{acro,appendix,booktabs}
  \usepackage[compat]{aaspatch}
  \let\oldincludegraphics\includegraphics
  \renewcommand*\includegraphics{\oldincludegraphics
    [width=\textwidth, height=.5\textheight, keepaspectratio]}
  \newcommand*\multicitedelim{; }
}{
  \usepackage{authblk,booktabs,hyperref}
  \usepackage{basic,physics,astro,draft}
  \usepackage[mode=biblatex, hyperref]{apjbib}
  \ExecuteBibliographyOptions{embedlinks}
  \addbibresource{torus.bib}
}
\makeatletter
\ifarxiv{
  \@ifpackageloaded{astro}{\let\ast@symit\textit}{}
}{
  \let\oldcaption\caption
  \AtBeginDocument{\renewcommand*\caption[1]{\oldcaption{%
    \setmathfont[Extension=.otf]{texgyretermes-math}\bsc@umoverride#1}
    \setmathfont[Extension=.otf]{texgyretermes-math}\bsc@umoverride}}
  \usepackage{etoolbox}
  \patchcmd\drf@abstract{\small}{\small
    \setmathfont[Extension=.otf]{texgyretermes-math}\bsc@umoverride}{}{}
  \patchcmd\drf@abstract{\bigskipamount}{\bigskipamount
    \setmathfont[Extension=.otf]{texgyretermes-math}\bsc@umoverride}{}{}
}
\makeatother
\input{torus.acr}
\newcommand*\nosup{^{}}
\newcommand*\momsrc{\vec S^\su m_\su}
\ifarxiv
  {\newcommand*\ergsrc[1]{S^{\mkern-2mu\su e}_{\mkern-3.5mu\su{#1}}}}
  {\newcommand*\ergsrc[1]{S^\su e_\su{#1}}}
\begin{document}

\title{Radiation-driven outflows from and radiative support in dusty tori of
  active galactic nuclei}
\author{Chi-Ho Chan}
\author{Julian~H. Krolik}
\affil{Department of Physics and Astronomy, Johns Hopkins University,
Baltimore, MD 21218, USA}
\date{October 29, 2015}

\shorttitle{Radiative outflow and support in AGN tori}
\shortauthors{Chan \& Krolik}
\pdftitle{Radiation-driven outflows from and radiative support in dusty tori of
  active galactic nuclei}
\pdfauthors{Chi-Ho Chan, Julian H. Krolik}

\begin{abstract}
\ifaastex\raggedright{}
Substantial evidence points to dusty, geometrically thick tori obscuring the
central engines of \acp{AGN}, but so far no mechanism satisfactorily explains
why cool dust in the torus remains in a puffy geometry. Near-Eddington \ac{IR}
and \ac{UV} luminosities coupled with high dust opacities at these frequencies
suggest that radiation pressure on dust can play a significant role in shaping
the torus. To explore the possible effects of radiation pressure, we perform
three-dimensional \aclp{RHD} simulations of an initially smooth torus. Our code
solves the hydrodynamics equations, the time-dependent multi--angle group
\ac{IR} \ac{RT} equation, and the time-independent \ac{UV} \ac{RT} equation. We
find a highly dynamic situation. \Ac{IR} radiation is anisotropic, leaving
primarily through the central hole. The torus inner surface exhibits a break in
axisymmetry under the influence of radiation and differential rotation;
clumping follows. In addition, \ac{UV} radiation pressure on dust launches a
strong wind along the inner surface; when scaled to realistic \ac{AGN}
parameters, this outflow travels at
$\num{\sim5000}\,(M/\SI{e7}{\solarmass})^{1/4}|[L_\su{UV}/(0.1\,L_\su
E)]^{1/4}\,\si{\kilo\meter\per\second}$ and carries
$\num{\sim0.1}\,(M/\SI{e7}{\solarmass})^{3/4}|[L_\su{UV}/(0.1\,L_\su
E)]^{3/4}\,\si{\solarmass\per\year}$, where $M$, $L_\su{UV}$, and $L_\su E$ are
the mass, \ac{UV} luminosity, and Eddington luminosity of the central object
respectively.
\end{abstract}
\acresetall\acuse{NGC}\acuse{IC}

\section{Introduction}

The discovery of reflected broad emission lines hidden in polarized light of
type\nobreakdash-2 \acp{AGN} \citep[e.g.,][]{1985ApJ...297..621A,
1990ApJ...355..456M} came as a revelation to \ac{AGN} research in that it can
only be reasonably explained by a geometrically and optically thick structure
surrounding the central source. Further observations established the properties
of the obscurer. The ratio of type\nobreakdash-2 to type\nobreakdash-1 objects
implies a high torus covering fraction, although the exact value of the ratio,
as well as its dependence on luminosity and redshift, is still under debate
\citep[e.g.,][]{2003ApJ...598..886U, 2010ApJ...714..561L, 2014MNRAS.437.3550M,
2015ApJ...806..127D, 2015ApJS..219....1O}. There is unequivocal proof for dust
\citep[e.g.,][]{1985aagq.conf..259M}; in particular, the broad
\SIrange{\sim1}{\sim100}{\micro\meter} bump in the \ac{SED} is attributed to
thermal radiation from warm dust \citep[e.g.,][]{1981ApJ...250...87R,
1987ApJ...320..537B, 1989ApJ...347...29S, 1993ApJ...418..673P}, and the cutoff
at \SI{\lesssim2}{\micro\meter} is indicative of dust close to sublimation
\citep[e.g.,][]{1969Natur.223..788R}. Finally, the existence of ionization
\citep[e.g.,][]{1989ApJ...345..730P, 1996VA.....40...63W} and scattering
\citep[e.g.,][]{1993ApJ...404..563P, 2005AJ....129.1212Z, 2006AJ....132.1496Z}
cones also signifies a small, geometrically and optically thick, toroidal
structure with an opening spanning a fraction of the solid angle around the
central source. \Ac{IR} interferometry has provided the first direct
observation of the obscuring torus in the form of warm dust within several
parsecs from the center in \ac{NGC}~1068 \citep{2004A&A...418L..39W,
2004Natur.429...47J, 2006A&A...450..483P, 2009MNRAS.394.1325R}, \ac{NGC}~4151
\citep{2003ApJ...596L.163S, 2009ApJ...705L..53B, 2010ApJ...715..736P},
Centaurus~A \citep{2007A&A...471..453M}, Circinus \citep{2007A&A...474..837T,
2012JPhCS.372a2035T, 2014A&A...563A..82T}, and other nearby \acp{AGN}
\citep{2008A&A...486L..17B, 2012ApJ...755..149H}. A sample of 29~\acp{AGN} have
thus far been studied in this way \citep{2009A&A...502...67T,
2009A&A...507L..57K, 2011A&A...527A.121K, 2011A&A...536A..78K,
2013ApJ...775L..36K, 2013A&A...558A.149B, 2016arXiv160205592L}. The
preponderance of evidence in favor of the torus inspires the idea that
observational variations between \ac{AGN} types~1 and~2 can be attributed to
orientation \citep[e.g.,][]{1989ApJ...336..606B, 1993ARA&A..31..473A,
1995PASP..107..803U}.

A crucial missing piece to this \ac{AGN} unification picture is an
understanding of torus dynamics. The torus has an aspect ratio of unity if its
velocity dispersion is comparable to its orbital velocity. If the velocity
dispersion were entirely due to thermal motion, hydrogen atoms at a distance
$r$ from \iac{SMBH} of mass $M$ would have temperature $\num{\gtrsim1.7e6}\,
(M/\SI{e7}{\solarmass})|(r/\si{\parsec})^{-1}\,\si{\kelvin}$, hot enough to
destroy dust by sputtering \citep{1988ApJ...329..702K}. Many models of
angle-dependent obscuration in \acp{AGN} have been put forward over the past
decades in an effort to solve this problem. They fall into five general
categories, but as we shall show below, none of them is entirely satisfactory.

Some proposed intrinsically warped structures. For example,
\ifaastex{\citet{1989tad..conf..457P} and \citet{1989ApJ...347...29S}}
{\citet{1989tad..conf..457P, 1989ApJ...347...29S}} advanced the notion that in
lieu of a torus, obscuration could be provided by a geometrically thin warped
disk. The disk must stretch from \SIrange{\sim1}{\sim e4}{\parsec} to reproduce
the observed \ac{IR} spectrum, at odds with the presence of well-defined
ionization cones on \SI{\sim100}{\parsec} scales, with \ac{IR} interferometric
observations, and with optical variability on a timescale of years
\citep[e.g.,][]{1989ApJ...340..190G}. Worse still, the covering fraction is
less than half except for the most severe warps and twists, and twists are
imperative if one must obstruct more than half of the lines of sight at high
inclinations. Parsec-scale warps and twists have garnered recent attention,
with proponents arguing that they can be sustained by stochastic accretion of
clumps from random directions \citep{2010ApJ...714..561L, 2012MNRAS.420..320H},
or that they are bending modes excited by radial flows caused by a lopsided
disk \citep{2012MNRAS.420..320H}. However, the torus advocated by
\citet{2010ApJ...714..561L} still suffers from the same shortcomings above,
whereas the aspect ratio of the \citet{2012MNRAS.420..320H} torus is only
\num{\sim0.1}.

Another option to partially avoid the dynamical problem is dust clumping.
Collisions between clumps can convert orbital shear to bulk velocity dispersion
\citep{1988ApJ...329..702K}. The collision rate must be almost once per orbit
for the mid-plane to be completely covered. Should these supersonic encounters
be inelastic, the resulting shocks would quickly turn the velocity dispersion
of clumps into internal energy; a torus that cools efficiently would settle to
the mid-plane, and one that does not would be geometrically thick, but so hot
that dust is burnt away. Clumps threaded with magnetic fields could be
sufficiently elastic, but the conditions are rather unusual, and one would ask
how adequate field strength could be sustained.

Other workers turn to large-scale magnetic fields for an answer. Dusty
molecular material lifted up from the surface of the accretion disk around the
central mass could be entrained in a magnetocentrifugal wind; in this scenario,
the torus is merely the parts of the wind which happen to be optically thick
enough \citep{1994ApJ...434..446K}. The dust perhaps takes the form of
optically thick clumps embedded in the wind
\citetext{\citealp{2006ApJ...648L.101E}\multicitedelim\citealp[see
also][]{1999ApJ...513..180K}}. Alternatively, magnetic fields could directly
support a static torus against gravity \citep{1998A&A...338..856L}. Magnetic
models, however, require strong, ordered fields on large scales, which are
difficult to justify.

Still another alternative is to invoke the nuclear starbursts seen in some
Seyfert~2s \citep[e.g.,][]{1997ApJ...482..114H, 2001ApJ...546..845G,
2001ApJ...558...81C, 2007ApJ...671.1388D}. They prompted
\citet{2002ApJ...566L..21W} to suggest turbulence stirred up by supernovae as a
means of creating a quasi-stable torus, but its size needs to be
\SI{\gtrsim30}{\parsec}, and even then the covering fraction is
\num{\lesssim0.2}. The obscuring gas disk of \citet{2016MNRAS.458..816H} has
similar drawbacks in that its size and aspect ratio are \SI{\gtrsim10}{\parsec}
and \num{\lesssim0.3}. Stellar feedback is in fact too weak to keep the torus
geometrically thick on parsec scales \citep{1988ApJ...329..702K}. Attacking the
problem from a different perspective, \citet{2009MNRAS.393..759S} considered
mass and energy injection by stars in a spherical and isotropic nuclear
cluster. Filaments in that scheme are formed by shock waves from supernovae and
planetary nebulae interacting with one another, while cold clumps come from
cooling. An analogous proposal by \citet{2013ApJ...766...92H} looked at
supernova ejecta and stellar winds released with some angular momentum. The gas
cools and is compressed to filaments, which then flows inward and accumulates
at the centrifugal barrier, forming a torus made geometrically thick by
X\nobreakdash-ray heating. Both models attempt to circumvent the weakness of
stellar feedback by injecting gas at the positions of the stars of a spatially
extended cluster, hence the fate of the torus is unclear once the starburst
ends. Moreover, the specific mass injection rate in the latter model is
\num{\sim6e3} times the galactic specific star formation rate.

Last but not least, \citet{1992ApJ...399L..23P} realized that since dust
opacity in the \ac{IR} is \num{\gtrsim10} times Thomson opacity, even
sub-Eddington \ac{AGN} luminosities could dramatically affect the torus through
radiation pressure. In their picture, \ac{UV} radiation from the central source
is converted to \ac{IR} on the inner surface of a smooth cylindrical torus
\citep{1992ApJ...401...99P}; part of the \ac{IR} radiation diffuses through the
torus and supports it. \Citet{2007ApJ...661...52K} revisited the problem and
constructed an analytic solution of a smooth axisymmetric torus under the
combined influence of gravity and radiation; \citet{2008ApJ...679.1018S} later
extended his work by incorporating the effects of hard X\nobreakdash-ray and
stellar heating. Unfortunately, both models are overly simplistic in that they
assume a hydrostatic torus and the diffusion approximation for the \ac{IR}
radiative flux.

Others have developed ideas along a similar vein. For example,
\citet{1999ApJ...521L..13O, 2001A&A...371..890O} considered radiation pressure
from both \iac{AGN} and a nuclear starburst ring, yet their obscuring structure
is stable near the mid-plane only for specific parameters.
\Citet{2012ApJ...749...32K} studied a magnetocentrifugal wind accelerated by
radiation from an accretion disk; the wind again depends on the existence of
some postulated large-scale magnetic field. An alternative model from
\citet{2012ApJ...758...66W, 2015ApJ...812...82W} focuses instead on turbulence
generated when gas streams lifted up by radiation fall back to the mid-plane
and intersect. Its conclusions can only be tentative because \ac{UV} heating
and radiative cooling in this model assume ionization by starlight while
X\nobreakdash-ray heating is based on stellar-mass black hole X\nobreakdash-ray
spectra, entirely ignoring \ac{AGN} radiation. The model also does not treat
dust destruction by sputtering at temperatures \SI{\gtrsim e5}{\kelvin}. In
addition, the omission of reprocessed \ac{IR} radiation in these three schemes
renders their applicability to optically thick tori doubtful. Less directly
related is the suggestion from \citet{2005ApJ...630..167T} that a starburst
disk with Eddington luminosity in the \ac{IR} possesses a tenuous, dusty, and
geometrically thick atmosphere.

In a series of articles, \citeauthor{2012ApJ...761...70D}
\citep{2011ApJ...741...29D, 2012ApJ...747....8D, 2012ApJ...761...70D,
2016ApJ...819..115D} investigated the effects of \ac{IR} radiation pressure on
dusty tori using simulations that couple hydrodynamics and radiation.
Encouragingly, they found that gas evolves naturally to a geometrically thick
obscuring wind. However, there are two limitations to this work. They neglected
momentum deposition from direct \ac{UV} illumination. More worrisome is their
use of the \ac{FLD} approximation, which can yield radiative fluxes in
completely wrong directions wherever the optical depth is comparable to or
smaller than unity. This problem is especially troubling when the dynamical
effect of radiation is important \citep[e.g.,][]{2012ApJS..199....9D}, as it is
here. \Citet{2012ApJ...759...36R} took the complementary direction of
performing Monte Carlo \ac{RT} on dusty gas and calculating the radiative
acceleration. The fact that they find accelerations exceeding gravity
emphasizes that hydrodynamics and \ac{RT} should be treated together.

We adopt a different approach in this article. Our program is to conduct a
series of numerical experiments designed to yield physical insight into each of
the most prominent mechanisms affecting torus dynamics; by adding mechanisms
one at a time, we hope to be able to distinguish their effects. Only toward the
end of this process will it be appropriate to draw specific relations between
our results and observable quantities. We begin in this article by presenting
three-dimensional, time-dependent \acp{RHD} simulations of a dusty torus that
experiences radiative acceleration on dust due to \ac{UV} radiation from the
central source and diffuse \ac{IR} radiation in the torus. Our simulations used
the finite-volume hydrodynamics code Athena \citep{2008ApJS..178..137S}
augmented by its time-dependent \ac{RT} module \citep{2014ApJS..213....7J} for
\ac{IR} radiation and a new long-characteristics \ac{RT} module for \ac{UV}
radiation (\cref{sec:UV radiative transfer}). The code simultaneously solves
the time-dependent hydrodynamics and \ac{RT} equations; most notably, it solves
the \ac{RT} equations on a large number of grid rays rather than adopting
\foreign{ad hoc} closure prescriptions. We leave other ingredients, such as
magnetic fields, realistic atomic and molecular heating and cooling rates, and
dust destruction by sputtering in high-temperature regions, to future
iterations.

In interpreting these results, it must be remembered that since the character
of the system demands mass loss from the inner surface, realistic tori
\emph{must} be resupplied externally. Our simulation, and indeed any other
simulation beginning with a finite amount of mass, \emph{cannot} portray
steady-state tori. The common device of putting a large gas reservoir at large
distances would impose a misleading radiative boundary condition. For this
reason, any connection between simulated and real tori must be posed in terms
of the rate of mass resupply necessary to secure stationarity.

We dedicate \cref{sec:methods} to our equations and simulation parameters.
Results are presented in \cref{sec:results}, while discussion can be found in
\cref{sec:discussion}.

\section{Methods}
\label{sec:methods}

We consider a cold, dusty, and optically thick torus orbiting a point mass $M$
at the origin. Isotropic \ac{UV} radiation of luminosity $L_\su{UV}$ emerges
from the origin. \Ac{UV} radiation impinging on the inner surface is absorbed
by dust and re-emitted in the \ac{IR}; radiation pressure from both the \ac{IR}
and the \ac{UV}, in concert with rotation, supports the torus against gravity.
Cylindrical coordinates $(R,\phi,z)$ are a natural choice for describing this
system, although we do occasionally refer to the spherical radius
$r\eqdef(R^2+z^2)^{1/2}$. From now on, the adjective `radial' shall implicitly
refer to the cylindrically radial direction. We also call the section of the
inner surface near the mid-plane the `inner edge.'

\subsection{Hydrodynamics}

We begin by examining the equations governing the dynamics of the torus. The
hydrodynamics equations are
\begin{align}
\label{eq:gas mass}
\pd\rho t+\divg(\rho|\vec v) &= 0, \\
\label{eq:gas momentum}
\pd{}t(\rho|\vec v)+\divg(\rho|\vec v|\vec v+p|\tsr I) &=
  -\rho|\grad\Phi+\momsrc{IR}+\momsrc{UV}, \\
\label{eq:gas energy}
\pd Et+\divg[(E+p)|\vec v] &=
  -\rho|\vec v\cdot\grad\Phi+\ergsrc{IR}+\ergsrc{UV}.
\end{align}
Here $\rho$, $\vec v$, and $p$ are gas density, velocity, and pressure. Gas
temperature and total energy density are $T=p/(\rho R_\su{ideal})$ and
$E=\ifaastex{\rho v^2/2}{\tfrac12|\rho v^2}+p/(\gamma-1)$, where $R_\su{ideal}$
and $\gamma$ are the specific ideal gas constant and the ratio of specific
heats. The gravitational potential of the central mass is $\Phi(\vec r)=-GM/r$.
The energy and momentum source terms due to radiation are $\ergsrc{IR,UV}$ and
$\momsrc{IR,UV}$; we shall define the \ac{IR} source terms in \cref{sec:IR
radiative transfer}, and the \ac{UV} source terms in \cref{sec:UV radiative
transfer}. Finally, the isotropic rank-two tensor is denoted by $\tsr I$.

The presence of dust means that gas temperature is \SI{\lesssim e5}{\kelvin},
otherwise dust would be rapidly destroyed by sputtering. This temperature is
much smaller than the virial temperature, hence the gas sound speed is also a
tiny fraction of the orbital velocity, or $c_\su s/v_\phi\ll1$. Because gas
pressure alone falls far short of maintaining the geometrical thickness of the
torus, it is dynamically unimportant compared to whatever pressure that
actually provides support against gravity, such as \ac{IR} radiation pressure,
so an approximate equation of state for the gas is entirely satisfactory. This
approximation breaks down outside the torus body, particularly in the central
hole where photoionization heating and Compton recoil can strongly heat the gas
\citep{1986ApJ...308L..55K, 2001ApJ...561..684K}. In the interest of focusing
attention on radiation-driven dynamics, in the simulations presented here we do
not change the equation of state between the body and the central hole. We plan
in future work to incorporate photoionization heating and related processes;
the increased gas pressure in the central hole could potentially alter the
shape of the inner surface.

We treat dust and gas as a single fluid with common velocity and temperature.
The fact that dust contributes significantly to \ac{IR} emission implies a dust
temperature below sublimation. We expect hydrogen at such temperature to remain
molecular and the vibrational modes of the molecule to be weakly excited; we
therefore set $R_\su{ideal}=\kB/(2|m_\element H)$ and
$\gamma=\ifaastex{7/5}{\tfrac75}$.

\subsection{\texorpdfstring{\Acl*{RT}}{Radiative transfer}}

Dust has \numrange{\sim e2}{\sim e3} times greater opacity to \ac{UV} radiation
than to \ac{IR} radiation \citep[e.g.,][]{2003A&A...410..611S}; such a large
contrast compels us to treat radiation at the two frequencies separately.

\Ac{UV} radiation comes from the innermost regions of an accretion disk at the
origin, but the angular distribution of its radiative flux is poorly known. The
classical picture of a limb-darkened disk only holds for a Newtonian,
scattering-dominated, geometrically thin, and optically thick disk; disk
turbulence, thermal instabilities, coronal scattering, as well as relativistic
boosting, beaming, and ray-bending, could all skew the angular profile of
emergent radiation. The axis of the disk also need not be aligned with that of
the torus. Because we lack a detailed disk model, and because our desire is to
understand physical principles rather than to provide observables, we simply
allow our \ac{UV} radiative flux to be isotropic instead of giving it a more
complicated and more model-dependent angular distribution.

Several \ac{RT} modules have already been developed for Athena. The
time-independent module \citep{2012ApJS..199....9D, 2012ApJS..199...14J}
performs \ac{RT} on a snapshot of the simulation, computes the Eddington
tensor, and uses it to close the angular moments of the \ac{RT} equation. In
comparison, the time-dependent module \citep{2014ApJS..213....7J} tracks the
propagation of radiation by solving the multi--angle group \ac{RT} equation
directly. Both modules are suited to handling diffusive \ac{IR} radiation
inside the torus, but we are restricted to the time-dependent module because it
is the only one available for cylindrical coordinates.

None of these modules is appropriate for point-source radiation crossing the
optically thin region between the central source and the torus because they
concentrate radiation along directions defined by the angle grid. Contours of
constant radiation energy density, instead of being spherically symmetric, show
prominent spherically radial spikes coincident with the grid rays. We therefore
reserve the time-dependent module for reprocessed \ac{IR} radiation inside the
torus. \Ac{UV} radiation emitted by the central source is handled with the
method of long characteristics, as described in \cref{sec:UV radiative
transfer}.

\subsubsection{Time-dependent \texorpdfstring{\acs*{IR} \acs*{RT}}{IR radiative
transfer}}
\label{sec:IR radiative transfer}

To first order in $v/c$, where $c$ is the speed of light, the mixed-frame
time-dependent \ac{RT} equation for \ac{IR} radiation interacting with gray
material reads \citep{2014ApJS..213....7J}
\begin{multline}\label{eq:radiative transfer}
\frac1c|\pd{I_\su{IR}}t+\uvec n\cdot\grad I_\su{IR}=
  \Bigl(-1+\uvec n\cdot\frac{\vec v}c\Bigr)|
  \rho|(\kappa_\su{IR}+\sigma_\su{IR})|I_\su{IR} \\
+\Bigl(1+3\,\uvec n\cdot\frac{\vec v}c\Bigr)|
  \rho|(\kappa_\su{IR}|B+\sigma_\su{IR}|J_\su{IR})
  -2|\rho|\sigma_\su{IR}|\frac{\vec v}c\cdot\vec H_\su{IR} \ifaastex{}\\
+\rho|(\kappa_\su{IR}-\sigma_\su{IR})|\frac{\vec v}c\cdot
  (\vec H^0_\su{IR}-\vec H_\su{IR}).
\end{multline}
The specific intensity integrated over the \ac{IR} in the observer frame is
$I_\su{IR}(\uvec n)$; its lowest three angular moments are $J_\su{IR}$, $\vec
H_\su{IR}$, and $\tsr K_\su{IR}$, from which the \ac{IR} radiation energy
density and flux follow as $E_\su{IR}=(4\pi/c)|J_\su{IR}$ and $\vec
F_\su{IR}=4\pi|\vec H_\su{IR}$. The frequency-integrated blackbody mean
intensity is $B(T)=c|\aSB|T^4/(4\pi)$, where $\aSB$ is the radiation constant.
The coupling between gas and radiation is mediated by $\kappa_\su{IR}$ and
$\sigma_\su{IR}$, the comoving absorption and scattering cross sections per
mass in the \ac{IR}. If we take the zeroth and first angular moments of
\cref{eq:radiative transfer}, we get
\ifaastex{
  \begin{alignat}{2}
  \label{eq:radiative zeroth moment}
  \frac1c|\pd{J_\su{IR}}t+\divg\vec H_\su{IR} &=
    \rho|\kappa_\su{IR}|(B-J_\su{IR})
    +\rho|(\kappa_\su{IR}-\sigma_\su{IR})|\frac{\vec v}c\cdot\vec H^0_\su{IR}
    &&\eqdef -\frac1{4\pi}|\ergsrc{IR}, \\
  \label{eq:radiative first moment}
  \frac1c|\pd{\vec H_\su{IR}}t+\divg\tsr K_\su{IR} &=
    \rho|\kappa_\su{IR}|\frac{\vec v}c|(B-J_\su{IR})
    -\rho|(\kappa_\su{IR}+\sigma_\su{IR})|\vec H^0_\su{IR}
    &&\eqdef -\frac c{4\pi}|\momsrc{IR}.
  \end{alignat}
}{
  \begin{alignat}{2}
  \nonumber
  & \frac1c|\pd{J_\su{IR}}t+\divg\vec H_\su{IR}= \\
  \label{eq:radiative zeroth moment}
  &\quad \rho|\kappa_\su{IR}|(B-J_\su{IR})
    +\rho|(\kappa_\su{IR}-\sigma_\su{IR})|\frac{\vec v}c\cdot\vec H^0_\su{IR}
    &&\eqdef -\frac1{4\pi}|\ergsrc{IR}, \\
  \nonumber
  & \frac1c|\pd{\vec H_\su{IR}}t+\divg\tsr K_\su{IR}= \\
  \label{eq:radiative first moment}
  &\quad \rho|\kappa_\su{IR}|\frac{\vec v}c|(B-J_\su{IR})
    -\rho|(\kappa_\su{IR}+\sigma_\su{IR})|\vec H^0_\su{IR}
    &&\eqdef -\frac c{4\pi}|\momsrc{IR}.
  \end{alignat}
}
The remaining piece to specify in \cref{eq:radiative transfer,eq:radiative
zeroth moment,eq:radiative first moment} is $\vec H^0_\su{IR}$, the first
angular moment of the \ac{IR} specific intensity in the fluid frame. It is
related to the angular moments in the observer frame by a Lorentz
transformation \citep{1984oup..book.....M}:
\begin{equation}
\vec H^0_\su{IR}=\vec H_\su{IR}
  -\frac{\vec v}c|J_\su{IR}-\frac{\vec v}c\cdot\tsr K_\su{IR}+\bigO(v^2/c^2).
\end{equation}

An unreasonably small time step is needed for the time-dependent \ac{RT} module
if $v\ll c$, but the fact that radiation relaxes to equilibrium much faster
than the hydrodynamic timescale means we can circumvent the problem with the
reduced speed of light approximation \citep{2001NewA....6..437G,
2013ApJS..206...21S}. The details are in \cref{sec:reduced light speed}; for
now, it suffices to know that the approximation replaces the physical light
speed $c$ attached to the time derivatives in \cref{eq:radiative
transfer,eq:radiative zeroth moment,eq:radiative first moment} with the reduced
light speed $\hat c$ subject to the requirement $v<\hat c\ll c$. An improvement
to how the time-dependent \ac{RT} module treats scattering is set forth in
\cref{sec:scattering solution}.

\subsubsection{\texorpdfstring{\acs*{IR}}{IR} and
\texorpdfstring{\acs*{UV}}{UV} opacities}
\label{sec:opacity}

The chief sources of opacity in our system are dust absorption and electron
scattering, which we model as
\begin{align}
\kappa_\su{IR}(T) &\eqdef \bar\kappa_\su{IR}\times
  \frac12|\left[1-\tanh\frac{\log_{10}(T/T_\su{ds})}{\Delta_\su{ds}}\right], \\
\kappa_\su{UV}(T) &\eqdef \bar\kappa_\su{UV}\times
  \frac12|\left[1-\tanh\frac{\log_{10}(T/T_\su{ds})}{\Delta_\su{ds}}\right], \\
\sigma_\su{IR}(T) &\eqdef \kappaT\times
  \frac12|\left[1+\tanh\frac{\log_{10}(T/T_\su{hi})}{\Delta_\su{hi}}\right].
\end{align}
In these fitting formulae, $T_\su{ds}\approx\SI{1500}{\kelvin}$ is the dust
sublimation temperature \citep[e.g.,][]{1969Natur.223..788R,
1981ApJ...250...87R, 1987ApJ...320..537B}, $T_\su{hi}\approx\SI{4013}{\kelvin}$
is the temperature at which hydrogen atoms in \ac{LTE} at a number density of
\SI{e4}{\per\cubic\centi\meter} are collisionally half-ionized, and
$\kappaT\approx\SI{0.397}{\centi\meter\squared\per\gram}$ is the Thomson
scattering cross section per mass. The dust opacities are normalized to Thomson
as $\bar\kappa_\su{IR}/\kappaT=20$ and $\bar\kappa_\su{UV}/\kappaT=80$, a
choice we shall justify in \cref{sec:realistic parameters}; the parameters
governing the transition between opacity regimes are $\Delta_\su{ds}=0.05$ and
$\Delta_\su{hi}\approx0.196$.

\subsection{Simulation setup}

We now spell out in detail the initial and boundary conditions, as well as
various tricks to keep the simulation stable.

\subsubsection{Initial condition}
\label{sec:initial condition}

The initial condition is based on the analytic solution of an axisymmetric
hydrostatic torus by \citet{2007ApJ...661...52K}. To summarize, the radiation
energy density inside the torus is determined along the mid-plane by
\ifaastex{
  \begin{equation}
  E^0_\su{IR}(R,0)\eqdef(E^0_\su{IR})_\su{in\vphantom0}\nosup
    +\frac{3GM|\rho_\su{in}}{R_\su{in}}|\left\{
    \frac1{1+\xi}|\biggl[\left(\frac R{R_\su{in}}\right)^{-(1+\xi)}-1\biggr]
    -\frac{j_\su{in}^2}\xi|
    \biggl[\left(\frac R{R_\su{in}}\right)^{-\xi}-1\biggr]\right\},
  \end{equation}
}{
  \begin{multline}
  E^0_\su{IR}(R,0)\eqdef(E^0_\su{IR})_\su{in\vphantom0}\nosup
    +\frac{3GM|\rho_\su{in}}{R_\su{in}}\times{} \\
  \left\{\frac1{1+\xi}|
    \biggl[\left(\frac R{R_\su{in}}\right)^{-(1+\xi)}-1\biggr]
    -\frac{j_\su{in}^2}\xi|
    \biggl[\left(\frac R{R_\su{in}}\right)^{-\xi}-1\biggr]\right\},
  \end{multline}
}
and everywhere else by the constant\nobreakdash-$E^0_\su{IR}$ contours
\begin{equation}
\frac12|\left(\frac z{R_\su{in}}\right)^2
  +\frac12|\left(\frac R{R_\su{in}}\right)^2
  -\frac13|j_\su{in}^2|\left(\frac R{R_\su{in}}\right)^3=\const.
\end{equation}
Of the five free parameters, four pertain to quantities measured at the inner
edge: radial coordinate $R_\su{in}$, gas density $\rho_\su{in}$, comoving
\ac{IR} radiation energy density $(E^0_\su{IR})_\su{in\vphantom0}\nosup$, and
ratio of gas orbital to Keplerian velocity $j_\su{in}$. The remaining free
parameter is the radial power-law exponent $\xi$ of gas density along the
mid-plane. We distinguish between $E_\su{IR}$ and $E^0_\su{IR}$, the \ac{IR}
radiation energy density in the observer and fluid frames respectively.
Although the radiative initial condition inside the torus is fully specified by
$E^0_\su{IR}$, the procedure for assigning $I_\su{IR}$ to individual grid rays
is somewhat elaborate, and is therefore relegated to \cref{sec:radiative
initial condition}. Gas density inside the torus is given by
\begin{equation}
\rho(R,z)\eqdef-\left[\frac{3GM}{R^2}|
  \left(1-j_\su{in}^2|\frac R{R_\su{in}}\right)\right]^{-1}|\pd{E^0_\su{IR}}R;
\end{equation}
in particular, $\rho(R,0)=\rho_\su{in}|(R/R_\su{in})^{-\xi}$. Gas temperature
and pressure are established by thermal equilibrium between gas and radiation,
to wit, $E^0_\su{IR}=\aSB|T^4$. Lastly, gas inside the torus has orbital
velocity
\begin{equation}\label{eq:IC velocity}
\vec v\eqdef j_\su{in}|\left(\frac{GM}{R_\su{in}}\right)^{1/2}\,\uvec e_\phi;
\end{equation}
in other words, $j/j_\su{in}=(R/R_\su{in})^{1/2}$, where $j\eqdef
v_\phi|(R/GM)^{1/2}$. This velocity profile in fact applies to all hydrostatic
radiation-supported tori in point-mass potentials (\cref{sec:velocity
profile}). The torus has extent $1<R/R_\su{in}<j_\su{in}^{-2}$ and
$z^2/R_\su{in}^2\le\ifaastex{}{\tfrac13|}(j_\su{in}^{-2}-1)^2|(2|j_\su{in}^2+1)\ifaastex{/3}{}$.

The free parameters are selected in a similar fashion to
\citet{2007ApJ...661...52K}. We pick $j_\su{in}=\ifaastex{1/2}{\tfrac12}$ such
that the inner edge is not supported by rotation alone, and that its vertical
extent is comparable to its radial coordinate. Having
\begin{equation}\label{eq:IC inner edge energy ratio}
(E^0_\su{IR})_\su{in\vphantom0}\nosup=\frac{GM|\rho_\su{in}}{R_\su{in}}
\end{equation}
ensures \ac{IR} and gravitational accelerations are comparable, that is,
$\norm{\grad E_\su{IR}}/\rho\sim GM/r^2$, although
$(E^0_\su{IR})_\su{in\vphantom0}\nosup$ could take on any value as long as
$E^0_\su{IR}\ge0$ inside the torus. The only deviation from
\citet{2007ApJ...661...52K} is in our choice that $\xi=1$, which results in a
less massive torus.

This initial condition is not an exact equilibrium since the central source may
not be able to maintain the initial distribution of \ac{IR} radiation energy
density along the inner surface. It is not even intended to resemble the
quasi-steady state of a realistic, axisymmetric, radiation-supported torus
since its properties, such as its radial and vertical extent, can be
arbitrarily altered by manipulating, say, the parameter $j_\su{in}$. The
initial condition is merely an approximate analytic solution of a hydrostatic
radiation-supported torus; as such, it is a convenient initial condition to
employ.

The exterior of the torus is filled with isothermal and hydrostatic ambient
material; we grant it nonzero orbital velocity because static ambient material
is found to be numerically unstable. To determine the properties of the ambient
material, we build upon the method used by \citet{1986MNRAS.221..339G} for
slender tori. The gravitational term of the force equation is clearly the
gradient of some scalar field; if we stipulate polytropic gas and a power-law
orbital velocity profile, then both pressure and centrifugal terms are
gradients as well, and the force equation becomes an easily solvable algebraic
equation. Unlike the solution of \citet{1986MNRAS.221..339G}, which is an
expansion around some $(R,z)$, our solution is exact. The density, pressure,
and velocity given by our method assuming a polytropic index of unity are
\begin{align}
\rho_\su{amb}(R,z) &\eqdef \bar\rho_\su{amb}\exp\biggl[
  \frac{GM}{r|(c_\su s^2)_{\smash{\su{amb}}}\nosup}
  +\frac{(v_\phi^2/c_\su s^2)_{\smash{\su{amb}}}\nosup}
    {2-2|q_{\smash{\su{amb}}}}\biggr], \\
p_\su{amb}(R,z) &\eqdef
  \rho_\su{amb}(R,z)\times(c_\su s^2)_{\smash{\su{amb}}}\nosup, \\
\vec v_\su{amb}(R) &\eqdef
  \left(\frac{GM}{R_\su{amb}}\right)^{1/2}|
  \biggl(\frac R{R_\su{amb}}\biggr)^{1-q_\su{amb}}\,\uvec e_\phi.
\end{align}
The center of the simulation domain $R_\su{amb}$ sets the length scale of the
ambient material, while the other parameters are
$\bar\rho_\su{amb}=\num{2e-8}\,\rho_\su{in}$, $(c_\su
s^2)_{\smash{\su{amb}}}\nosup=GM/R_\su{amb}$, and $q_\su{amb}=1.75$. The shear
parameter must satisfy $1.5<q_\su{amb}<2$ in order that the ambient material
have finite height and be stable. Since it is preferable that density and
pressure vary monotonically across the torus boundary, we additionally require
$\rho\ge\rho_\su{amb}$ and $p\ge p_\su{amb}$ everywhere in the initial
condition.

\subsubsection{Central mass and reduced speed of light}
\label{sec:central mass}

The astute reader will notice that we have evaded any mention of the value of
the central mass $M$. This is because its choice is by far the most complicated
consideration in our simulations.

Sharp discontinuities in numerical calculations are flanked by ringing
artifacts, which resemble wiggles associated with the Gibbs phenomenon. These
artifacts usually damp out over time; however, in the case of a cylindrical
discontinuity in a gas partially supported against gravity, such as the inner
edge, the artifact grows rapidly at any spatial resolution. Experimentation
with different values of $c_\su s/v_\phi$ shows that the artifact can be
suppressed by demanding $c_\su s/v_\phi\gtrsim\bigO(0.1)$. If $c_\su s/v_\phi$
is kept at the low end of the numerically permitted range, gas pressure should
always be weak compared to gravity; as long as gas pressure is a minor
influence, it should not matter if it is not as tiny as in realistic
astrophysical circumstances.

These constraints determine $M$. The gas equilibrium temperature $T_\su{in}$ at
the inner edge is set by $\kappa_\su{UV}|L_\su{UV}/(4\pi|R_\su{in}^2)=
\kappa_\su{IR}|c|\aSB|T_\su{in}^4$; the stability requirement
$R_\su{ideal}|T_\su{in}/(GM/R_\su{in})\gtrsim\bigO(0.1)^2$ then becomes
\begin{align}
\nonumber
M &\lesssim \frac{R_\su{ideal}^2|(\kappa_\su{UV}/\kappa_\su{IR})}
  {G|\kappaT|\aSB|T_\su{ds}^2}|\left(\frac{L_\su{UV}}{L_\su E}\right)|
  \left(\frac{T_\su{in}}{T_\su{ds}}\right)^{-2}\bigO(0.1)^{-4} \\
&\approx \num{7.58e-4}\,\left(\frac{L_\su{UV}/L_\su E}{0.1}\right)|
  \left(\frac{T_\su{in}}{T_\su{ds}}\right)^{-2}\bigO(0.1)^{-4}\,
  \si{\solarmass},
\end{align}
with $L_\su E$ being the Eddington luminosity. We use
$M\approx\SI{0.758}{\solarmass}$ in practice. We shall argue in
\cref{sec:normalization} that our failure to simulate a torus around a genuine
supermassive black hole is completely superficial.

We now consider how $M$ affects our choice of $\hat c$. The dynamical timescale
is $[R_\su{in}^3/(GM)]^{1/2}$, whereas the \ac{IR} radiation diffusion
timescale in the reduced speed of light approximation is
$\rho_\su{in}|\bar\kappa_\su{IR}|[\ifaastex{}{\tfrac12|}(j_\su{in}^{-2}-1)|R_\su{in}\ifaastex{/2}{}]^2/\hat
c$. Clean separation of dynamical evolution from \ac{IR} radiation diffusion
requires
\begin{equation}
\frac{\hat c}{(GM/R_\su{in})^{1/2}}\gg
  \rho_\su{in}|\bar\kappa_\su{IR}|R_\su{in}\times\frac14|(j_\su{in}^{-2}-1)^2;
\end{equation}
the right-hand side is an overestimate by a factor of a couple because density
falls off away from the inner edge. We settle on $\hat
c\sim50\,(GM/R_\su{in})^{1/2}$ as a trade-off between accuracy and
computational time (see \cref{sec:normalization} for the actual value),
although we find little qualitative difference even at $\hat
c\approx8.94\,(GM/R_\su{in})^{1/2}$ as long as $v<\hat c$ everywhere.

\subsubsection{Normalization and parameters}
\label{sec:normalization}

Physical quantities are hereafter normalized to their respective fiducial
values. The fundamental fiducial quantities are the central mass $M$, the dust
sublimation temperature $T_\su{ds}$, and the Thomson scattering cross section
per mass $\kappaT$; all other fiducial quantities, listed in
\cref{tab:fiducial}, are derived from them. In particular, $L_\su E$ is the
Eddington luminosity, and $r_0$ is the distance where the effective temperature
of the radiative flux in vacuum from a source with Eddington luminosity equals
$\smash{\sqrt2}$ times the dust sublimation temperature. Note that a system in
which rotational support is provided by diffusive radiation must have
$\rho_0|v_0^2/r_0\sim E_0/r_0$.

\begin{table}
\caption{Derived fiducial quantities.}\label{tab:fiducial}
\begin{tabular}{ccc}
\toprule
Fiducial quantity & Symbol & Definition \\\midrule
luminosity & $L_\su E$ & $4\pi GMc/\kappaT$ \\
length & $r_0$ & $[L_\su E/(4\pi c|\aSB|T_\su{ds}^4)]^{1/2}$ \\
velocity & $v_0$ & $(GM/r_0)^{1/2}$ \\
time & $t_0$ & $(GM/r_0^3)^{-1/2}$ \\
gas density & $\rho_0$ & $(\kappaT|r_0)^{-1}$ \\
gas pressure & $p_0$ & $\rho_0|v_0^2=\aSB|T_\su{ds}^4$ \\
radiation energy density & $E_0$ & $L_\su E/(4\pi|r_0^2|c)=p_0$ \\
radiative flux & $F_0$ & $c|E_0$ \\
\bottomrule
\end{tabular}
\end{table}

One virtue of our normalization is that, because the characteristic length
scale is $r_0\propto M^{1/2}$, the gravitational acceleration at $r=r_0$ does
not depend on $M$. We can guarantee accelerations due to gas pressure and
radiation are likewise independent of $M$ by fixing $c_\su s/v_\phi$ and
$L_\su{UV}/L_\su E$ for each simulation. These invariances ensure that the
character of the dynamics simulated differs from that for more astrophysically
relevant values of $M$ only in the magnitude of the timescale $t_0\propto
M^{1/4}$. The normalizations of other quantities, such as momentum density,
could nevertheless vary with $M$.

Now that we have a system of normalization in place, we can translate our
choice $M\approx\SI{0.758}{\solarmass}$ in \cref{sec:central mass} to
dimensionless parameters that the simulation actually accepts, namely,
$R_\su{ideal}=0.05\,p_0/(\rho_0|T_\su{ds})$ and $c\approx\num{2.70e4}\,v_0$.

It remains to pick the appropriate parameters for the simulations. To start
with, we choose $0.10\le L_\su{UV}/L_\su E\le0.15$ in steps of 0.01 because
these luminosities are high enough to hold back the infall of the torus, but
low enough not to push it away too briskly. The simulation at each $L_\su{UV}$
is run for about two orbits at the inner edge, at which point the radial
component of velocity is positive throughout the torus body.

Three of the five parameters governing the initial condition have already been
picked in \cref{sec:initial condition}; the remaining two will be given here.
The inner edge $R_\su{in}$ should be just outside the dust sublimation radius
\citep[e.g.][]{1969Natur.223..788R, 1981ApJ...250...87R, 1987ApJ...320..537B,
1989ApJ...337..236C, 1989ApJ...347...29S, 1993ApJ...418..673P}, that is,
$R_\su{in}^2\gtrsim
r_\su{ds}^2=\kappa_\su{UV}|L_\su{UV}/(4\pi|\kappa_\su{IR}|c|\aSB|T_\su{ds}^4)$.
Our initial condition puts $R_\su{in}=0.8\,r_0$, so that $R_\su{in}$ goes from
$1.26\,r_\su{ds}$ to $1.03\,r_\su{ds}$ as $L_\su{UV}/L_\su E$ varies from 0.10
to 0.15. The reduced light speed introduced in \cref{sec:central mass} can be
recast in terms of fiducial values as $\hat c=50\,v_0$.

The density at the inner edge is selected to be $\rho_\su{in}=\rho_0$. The
radial Thomson optical depth of our initial condition along the mid-plane is
\begin{equation}
\int_{R_\su{in}}^\infty dR\,\rho|\kappaT=
  \rho_\su{in}|R_\su{in}|\kappaT\times
  \begin{cases}
    [j_\su{in}^{-2|(1-\xi)}-1]/(1-\xi), & \xi\ne1, \\
    2\ln j_\su{in}^{-1}, & \xi=1,
  \end{cases}
\end{equation}
while the vertical Thomson optical depth at $R=R_\su{in}$ is
\ifaastex{
  \begin{equation}
  \int_{-\infty}^\infty dz\,\rho|\kappaT=
    2|\rho_\su{in}|R_\su{in}|\kappaT
    \int_1^{j_\su{in}^{-2}}dx\,\frac{x^{-(2+\xi)}|(1-j_\su{in}^2|x)}
    {(x^2-\tfrac23|j_\su{in}^2|x^3-1+\tfrac23|j_\su{in}^2)^{1/2}}.
  \end{equation}
}{
  \begin{multline}
  \int_{-\infty}^\infty dz\,\rho|\kappaT= \\
  2|\rho_\su{in}|R_\su{in}|\kappaT
    \int_1^{j_\su{in}^{-2}}dx\,\frac{x^{-(2+\xi)}|(1-j_\su{in}^2|x)}
    {(x^2-\tfrac23|j_\su{in}^2|x^3-1+\tfrac23|j_\su{in}^2)^{1/2}}.
  \end{multline}
}
Our parameters yield Thomson optical depths of \num{\approx1.11} and
\num{\approx1.01} respectively, consistent with the observed range of values
\citep[e.g.,][]{1999ApJ...522..157R}. The corresponding \ac{IR} optical depths
are established by numerical integration to be \num{\approx19.9} and
\num{\approx10.9}. The ratios of Thomson to \ac{IR} optical depths are not
$\kappaT/\bar\kappa_\su{IR}$ due to the higher temperature and lower \ac{IR}
opacity near the inner edge.

The simulation domain spans
$[0.3\,r_0,5\,r_0]\times\ifaastex{[-\pi/4,\pi/4]}{[-\tfrac14|\pi,\frac14|\pi]}\times[-4\,r_0,4\,r_0]$
in $(R,\phi,z)$. The vertical direction is made as tall as possible to capture
escaping material, while not so tall that the centrifugal barrier would cause
numerical problems at the inner-radial boundary. The number of grid cells is
$188\times33\times320$ in $(R,\phi,z)$, and the number of grid rays per cell is
168.

\subsubsection{Boundary conditions and numerical limits}

Periodic hydrodynamic and radiative boundary conditions are adopted for the
azimuthal direction, with the understanding that grid rays at one boundary must
be rotated through $\pm\ifaastex{\pi/2}{\tfrac12|\pi}$ before they can be
copied to the ghost zones at the opposite boundary, to account for the fact
that the simulation domain covers only a quarter of a circle
(\cref{sec:normalization}).

Outflow hydrodynamic boundary conditions are applied at both boundaries in the
radial and vertical directions. The value of $\vec v$ in the ghost zones is
duplicated from the last physical cell, and components pointing into the
simulation domain are zeroed. We then adjust $\rho$ and $p$ in the ghost zones
at constant $c_\su s^2\eqdef p/\rho$ so that the pressure gradient exactly
cancels the gravitational and centrifugal forces. The value of $c_\su s^2$ is
the greater of $p/\rho$ of the last physical cell and $(c_\su
s^2)_{\smash{\su{amb}}}\nosup$; bounding $c_\su s^2$ from below protects $\rho$
and $p$ from numerical underflow.

Outflow radiative boundary conditions are used for the outer-radial and both
vertical boundaries. For grid rays pointing away from the simulation domain, we
copy their values of specific intensity from the last physical cell to the
ghost zones; for all other grid rays, we set their values of specific intensity
to zero. We also implement a cutout boundary condition for the inner-radial
boundary. Ghost zones are filled out in the same way as an outflow boundary; on
top of that, for every radially inward grid ray intersecting this boundary, we
trace its trajectory across the cylindrical cutout to where it re-enters the
domain, and add the specific intensity of the exiting grid ray to the
corresponding grid ray in the ghost zone at the re-entry point without allowing
for any time delay. Since the angle grid does not vary with coordinates, the
matching of exiting to re-entering grid rays is exact. Grid rays re-entering at
azimuthal coordinates outside the simulation domain are wrapped back after a
suitable rotation.

Limits on gas density and temperature are enforced for the sake of numerical
stability. We require that $\rho$ satisfy $\rho\ge\rho_\su{amb}$, and that $T$
satisfy $\num{e-3}\,T_\su{ds}\le T\le 10\,(c_\su
s^2)_{\smash{\su{amb}}}\nosup/R_\su{ideal}$; if at any time $\rho$ and $T$
violate these conditions, we reset them to the nearest value within the
acceptable range. The density floor guarantees a stable vacuum. A static
pressure floor is unsatisfactory because pressure could hit the floor before
density; any further drop in density would result in erroneous heating of the
gas, making the overall time step unreasonably small. A better approach is to
restrict temperature to within a generous range. Because
$\kappa_\su{IR,UV}\approx0$ when $T\gg T_\su{ds}$ (\cref{sec:opacity}) and
$R_\su{ideal}|T_\su{ds}\ll(c_\su s^2)_{\smash{\su{amb}}}\nosup$ if the torus is
supported by radiation pressure, radiative heating cannot bring the gas to the
temperature ceiling.

\section{Results}
\label{sec:results}

We now present the results of our simulations. It is important to remember that
the simulations do not reach steady state; therefore, what we report below is
the transient response of the \citet{2007ApJ...661...52K} hydrostatic torus to
\ac{UV} irradiation. We already see within the first two orbits that the torus
will never be hydrostatic for any $L_\su{UV}$. This is because the degree of
radiative support varies strongly with time and location (\cref{sec:gas
overview}), because the inner surface is corrugated radially by radiation and
sheared azimuthally by differential rotation (\cref{sec:inner surface
perturbation}), and because mass is continually lost in the form of a
radiation-driven wind from the inner surface (\cref{sec:gas overview,sec:loss
rates}). We do not claim our simulations represent the only possible
configuration of a torus; instead, we wish to draw qualitative conclusions that
apply to any flavor of smooth, radiation-supported torus, and to let this
information guide us toward constructing a more realistic torus.

\subsection{Qualitative description of gas motion}
\label{sec:gas overview}

For the purpose of orientation, we begin by examining the evolution of the
torus in general terms, using \cref{fig:time sequence} as reference. Parts of
the simulation domain with $z<0$ are discarded from our figures on the grounds
that we observe no breaking of symmetry about the mid-plane.

\begin{figure*}
\includegraphics{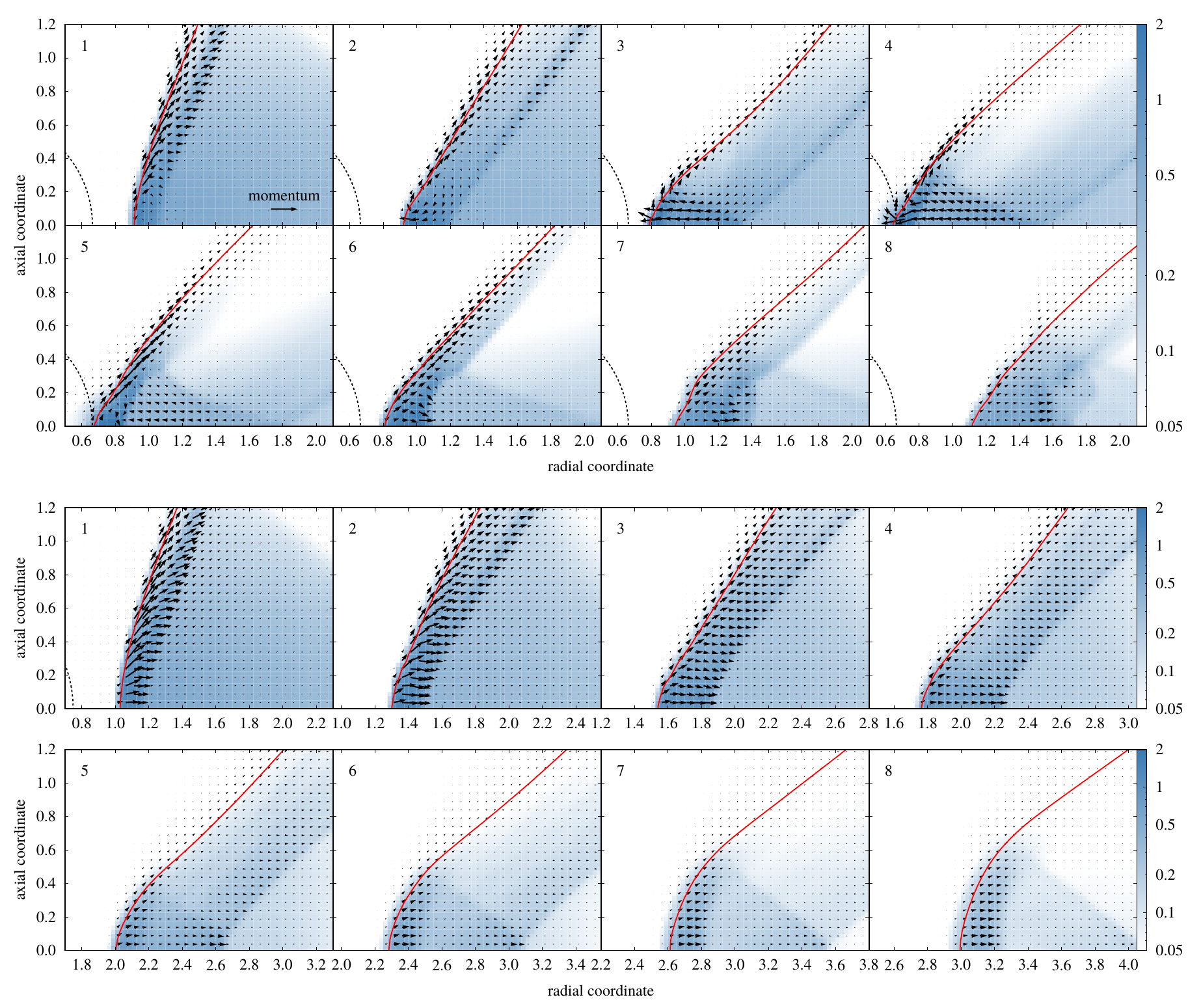}
\caption{Zoom-in of the azimuthally averaged poloidal plane at times $t=n|t_0$,
where $n$ is the number in the top-left corner of each panel. Gas density is
presented on a logarithmic scale as blue intensities (see color bar along the
right edge). The dust sublimation surface $r=r_\su{ds}$
(\cref{sec:normalization}) is the dashed black curve around the origin, and the
red contour traces the surface on which $\tau_\su{UV}=1$. Momentum density is
shown by arrows with lengths $\mathrelp\propto\rho v$; the arrow in the
bottom-right corner of the first panel has length $0.5\,\rho_0|v_0$. All
quantities are normalized to fiducial units (\cref{sec:normalization}).
\textit{Top grid:} Plot of the $L_\su{UV}/L_\su E=0.11$ simulation.
\textit{Bottom grid:} Plot of the $L_\su{UV}/L_\su E=0.14$ simulation; note
that the abscissa shifts at a constant rate toward the right from panel to
panel.}
\label{fig:time sequence}
\end{figure*}

\Ac{UV} radiation creates two immediate effects on the inner surface. Gas at
the inner surface is swept up in the radially outward direction, forming a
density concentration along it. The inner surface recedes as a result, first
supersonically, then subsonically; this excites a transient in the form of an
acoustic density perturbation peeling away from the density concentration and
propagating outward through the torus, discernible at times $t\gtrsim1$. The
perturbation is shaped like a chevron bending outward when viewed with the full
range of $z$.

More notably, \ac{UV} radiation shaves off gas at high latitudes and creates a
wind, while the central hole opens up from a cylindrical to a flaring shape.
There are two reasons why this gas is the most vulnerable to \ac{UV} stripping.
First, we designate the \ac{UV} optical depth from the central source by
$\tau_\su{UV}$. Only gas at $\tau_\su{UV}\lesssim1$ experiences substantial
\ac{UV} acceleration, and the $\tau_\su{UV}=1$ surface slants radially outward
with increasing $\abs z$ in the initial condition since $\rho$ diminishes
monotonically with $\abs z$. Second, let us mentally divide the solid angle as
seen from the origin into infinitely many sectors, and let us study the
dynamics of the gas column contained within each sector with the proviso that
neighboring columns do not interact. This is akin to the approach used by
\citet{2012ApJ...759...36R} to calculate accelerations in their simulations.
The acceleration of a column of thickness $\Delta r$ due to point-source
\ac{UV} radiation is $\mathrelp\propto(1-e^{-\rho|\kappa_\su{UV}|\Delta
r})/(\rho|\Delta r)$, an expression that drops with increasing $\rho|\Delta r$,
while for any plausible initial condition of the torus, including ours
(\cref{sec:initial condition}), $\rho|\Delta r$ rises with inclination, defined
as the angle from the polar axis. An intuitive way to think about the second
argument is that \ac{UV} radiative flux is spherically symmetric, and if it is
capable of supporting an optically thick column against gravity at low
latitudes, then it is fully equipped to expel an optically thin column at high
latitudes.

One might think that only gas at high latitudes participates in the wind, while
gas at low latitudes accelerated by \ac{UV} radiation is simply rammed against
the inner surface. This is untrue because gas pressure along the flaring inner
surface is virtually constant. At any height above the mid-plane, \ac{UV}
acceleration has a component parallel to the inner surface; unchecked by
pressure gradients, this component is free to peel off gas into a wind gliding
outward along the inner surface. Also note that, according to \cref{eq:IC
velocity}, gas starting out from smaller $R$ has smaller $R|v_\phi$, so the
wind preferentially removes gas with lower specific angular momentum.

Care must be exercised in reading \cref{fig:time sequence} after this initial
phase. We shall see in \cref{sec:inner surface perturbation} that the initially
axisymmetric inner surface becomes radially corrugated at $t\gtrsim4\,t_0$. For
$L_\su{UV}/L_\su E=0.11$, averaging this undulating structure in the azimuthal
direction produces the illusion that the inner surface at $t\gtrsim6\,t_0$
resembles a thick shell while in fact the density concentration remains thin in
any single poloidal slice. For $L_\su{UV}/L_\su E=0.14$, the inner surface
stays relatively axisymmetric; however, the fact that it moves radially outward
almost as quickly as the transients excited along it gives it the appearance of
multiple shells.

The radial motion subsequent to the initial phase depends on $L_\su{UV}$, which
determines the \ac{IR} radiative flux across the torus. For $L_\su{UV}/L_\su
E\ge0.13$, \ac{IR} radiative flux is strong enough that gas velocity is
radially outward in the torus body almost all the time, hence there is little
doubt the torus will be driven outward. In contrast, for $L_\su{UV}/L_\su
E\le0.12$, a region develops above and below the mid-plane at greater radial
coordinates than the inner edge in which the sum of the radial components of
\ac{IR} and centrifugal accelerations falls slightly short of counteracting
gravity, and thus the radial component of velocity is negative. The size of
this region decreases with $L_\su{UV}$. Gas outside the region continues to be
propelled outward, but gas inside slides slowly toward the mid-plane and
inward; as it reaches the $\tau_\su{UV}=1$ surface, it is flung away by \ac{UV}
radiation. This kind of inflow--outflow is essentially a balance between the
infall of gas toward the inner edge and the ability of \ac{UV} radiation to
clear out the pileup. Because there is only a finite amount of gas in the
simulated torus, the inflow--outflow in our simulations cannot last forever.

The density distribution at times $t\gtrsim4\,t_0$ bears little resemblance to
the initial condition. Gas continues to be removed in the wind, but the
detailed shape of the body depends on whether vertical support due to \ac{IR}
radiation is stronger or weaker than gravity. For $L_\su{UV}/L_\su E\ge0.14$,
\ac{IR} radiative flux is sufficiently strong to inflate the body in the
vertical direction. But for $L_\su{UV}/L_\su E\le0.13$, the body falls toward
the mid-plane, reaching a thickness comparable to the gas pressure scale
height, and then expands back vertically. The density concentration along the
inner surface is shaped like another chevron and is taller than the body thanks
to \ac{UV} radiation constantly accelerating the gas upward and outward.
Although the \ac{IR} covering fraction drops steadily with time, the vertically
extended inner surface and the wind keep it at a value higher than would be due
to the body alone.

The degree of \ac{IR} radiative support differs from place to place at these
late times. For all $L_\su{UV}$, \ac{IR} vertical support in the chevron-shaped
inner surface is generally insufficient to counteract gravity; as we move
radially outward, we encounter a wedge-shaped, lower-density region in which
marginal \ac{IR} vertical support prevails, followed by another region of even
lower density in which \ac{IR} vertical support again falls short of gravity.
As $L_\su{UV}$ increases, \ac{IR} vertical support becomes stronger more
rapidly at the inner surface than further outward in the torus, such that the
inner surface is completely supported against gravity at $L_\su{UV}/L_\su
E=0.15$ even when other parts of the torus are not.

Significant mass loss in the wind leads to a substantial drop in radial \ac{IR}
optical depth along the mid-plane over time: By $t=10\,t_0$, the optical depth
is less than half its initial value for $L_\su{UV}/L_\su E=0.10$, and down to
\num{\sim0.05} times its initial value for $L_\su{UV}/L_\su E=0.15$. This
diminution in optical depth can be quite uneven as a function of azimuthal
coordinate for $L_\su{UV}/L_\su E$ at the low end of the simulated range
because, as we shall discuss in \cref{sec:inner surface perturbation}, those
are the conditions in which the non-axisymmetric radial perturbation at the
inner surface grows the most; at the high end of $L_\su{UV}/L_\su E$,
axisymmetry is maintained much more closely.

The rate at which \ac{UV} radiation deposits momentum in the torus is
proportional to the \ac{UV} covering fraction, whereas the mass of the torus is
roughly proportional to the covering fraction times the optical depth; hence,
the sharp plunge in \ac{IR} optical depth explains why the body experiences
progressively stronger radially outward acceleration. For $L_\su{UV}/L_\su
E\le0.12$, this means the inflow--outflow eventually ceases, the radial
component of velocity turns positive throughout the body, and the body slides
outward more and more quickly as further mass loss accompanies its outward
motion.

\subsection{Radial perturbation of the torus inner surface}
\label{sec:inner surface perturbation}

Another intriguing complication at $t\gtrsim4\,t_0$ is the breaking of
axisymmetry along the inner surface.

The three-dimensional structure of the inner surface stays remarkably vertical
throughout the simulation for all $L_\su{UV}$. Isosurfaces of constant density
extend almost perpendicularly upward and downward from the mid-plane until they
are cut off at some height. This height depends on the density at the
isosurface, and typically increases with radial coordinate due to the flaring
shape of the inner surface (\cref{sec:gas overview}).

The verticality of the inner surface allows us to focus our attention on the
mid-plane, as we do in \cref{fig:RTI}. Non-axisymmetry along the mid-plane
assumes the form of a slight radial perturbation of the inner edge going
through three oscillations per quarter circle at $t\sim4\,t_0$. The
perturbation grows in amplitude afterward, and its behavior in the nonlinear
regime depends on $L_\su{UV}$.

\begin{figure*}
\includegraphics{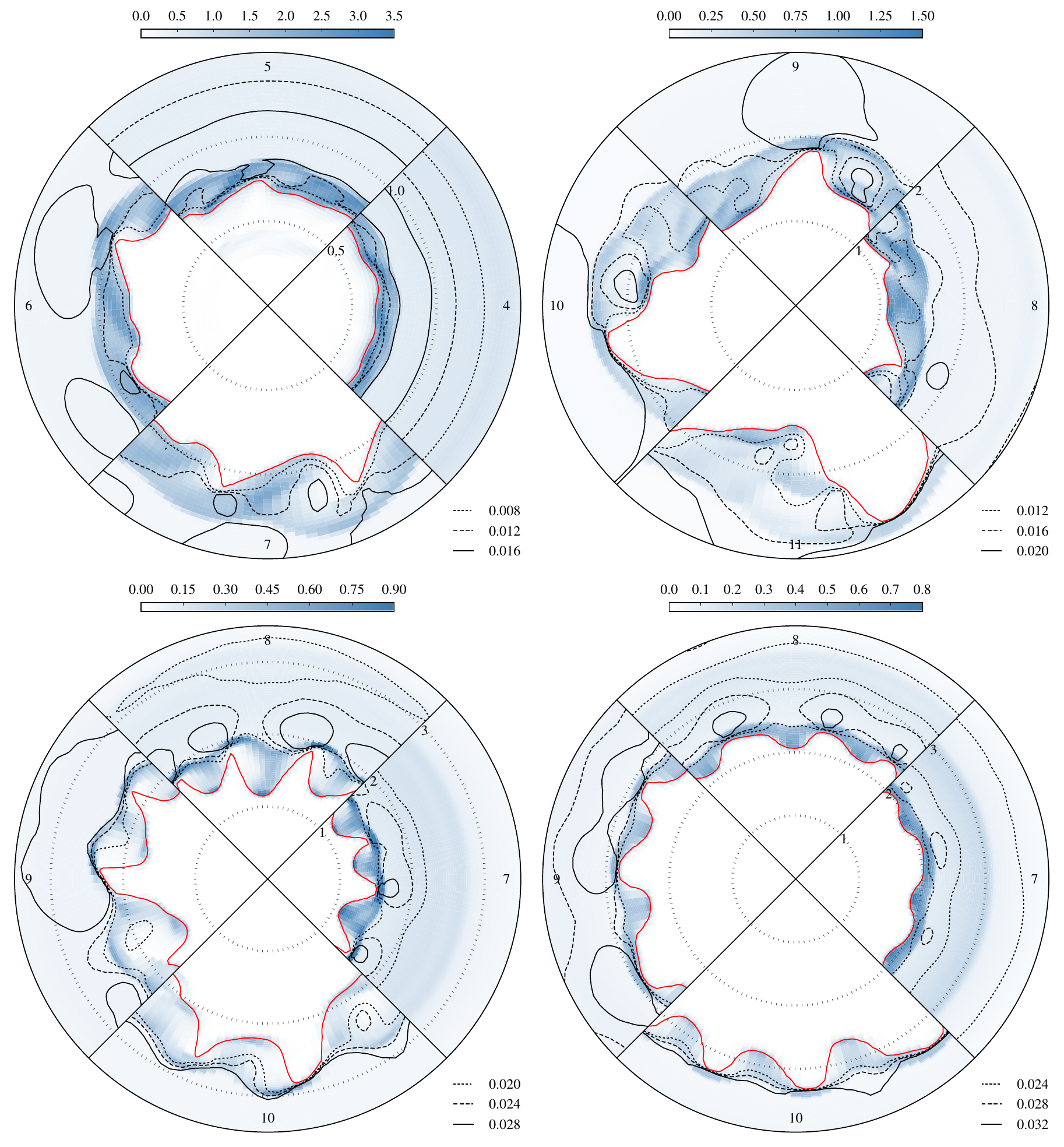}
\caption{Mid-plane at times $t=n|t_0$, where $n$ is the number along the rim of
each quadrant of the circles. Gas density is presented on a linear scale as
blue intensities (see color bar above each circle), the red contour traces the
surface on which $\tau_\su{UV}=1$, and the black contours display $R^2\,\uvec
e_R\cdot\vec F_\su{IR}$ at levels indicated in the legend. Orbital motion is
counter-clockwise, all quantities are normalized to fiducial units
(\cref{sec:normalization}), and all circles have different scales. \textit{Top
circles:} Plots of the $L_\su{UV}/L_\su E=0.11$ simulation. \textit{Bottom-left
circle:} Plot of the $L_\su{UV}/L_\su E=0.12$ simulation. \textit{Bottom-right
circle:} Plot of the $L_\su{UV}/L_\su E=0.13$ simulation.}
\label{fig:RTI}
\end{figure*}

The top circles illustrate how, for $L_\su{UV}/L_\su E\le0.11$, the originally
smooth inner surface breaks up into dense, thin sheets overlapping in the
azimuthal direction; seen along the mid-plane, the dominant sheets resemble
trailing spiral density waves. For $L_\su{UV}/L_\su E=0.11$, the three original
oscillations combine into one at $t\sim9\,t_0$; for $L_\su{UV}/L_\su E=0.10$,
the three oscillations merge into two at $t\sim5\,t_0$ and break apart into
three again at $t\sim12\,t_0$.

In comparison, the bottom circles show that for $L_\su{UV}/L_\su E\ge0.12$, the
inner edge is characterized by a series of fingers pointing radially inward,
connected at the outward end by arcs which are convex outward. The fingers are
better described in three dimensions as vertical inward protrusions of the
inner surface shaped like rounded flaps in poloidal section. The tips of the
fingers and the middle portions of the arcs are slightly denser than other
parts of the inner edge. The tips of the fingers are also sheared azimuthally
into hooks by differential rotation. At any given time, the amplitude of the
perturbation, as well as the azimuthal distortion of the fingers due to
shearing, both decrease with $L_\su{UV}$.

There is nothing physical about the number three in the number of oscillations
at $t\sim 4\,t_0$. The initial perturbation is seeded by a small numerical
artifact associated with the angle grid whose influence is the strongest at six
azimuthal coordinates; the six originally tiny oscillations then merge to three
easily discernible ones. Since the artifact is fixed in space while the orbital
motion of the gas takes it across azimuthal coordinates, the artifact is not
expected to act on the same gas packet continually; therefore, we believe the
growing perturbation is a real effect.

\subsection{Anisotropy of \texorpdfstring{\acs*{IR}}{IR} radiation}
\label{sec:IR overview}

We now discuss the properties of \ac{IR} radiation with the aid of
\cref{fig:overview}. Although the figure pertains to one snapshot of a single
simulation, it is representative of the configuration of the torus at earlier
times for all $L_\su{UV}$.

The first thing we notice in the top panel is that gas and \ac{IR} radiation
temperature contours coincide in the torus body, and diverge only in
low-density regions outside the body. This confirms our expectation that
thermal equilibrium holds deep inside the torus but not outside.

A more significant observation, verifiable by a quick inspection of the bottom
panel, is that \ac{IR} radiative flux streaming vertically through the central
hole is stronger by a factor of a few than its nearly horizontal counterpart
diffusing through the torus. This is explained by the conversion of \ac{UV}
radiation to \ac{IR} taking place in a thin layer of thickness
$\mathrelp\sim(\rho_\su{in}|\kappa_\su{UV})^{-1}$ centered at $\tau_\su{UV}=1$.
The \ac{IR} optical depth is \num{\gg1} from there to the outer surface, but
merely $\mathrelp\sim\kappa_\su{IR}/\kappa_\su{UV}\ll1$ to the central hole;
consequently, it is much easier for the freshly created \ac{IR} radiation to
head back into the central hole than to penetrate the body.

In a geometrically and optically thick torus, some of the \ac{IR} radiation
emitted by the inner edge can cross the central hole, reach the far side, and
be absorbed again, giving \ac{IR} radiation multiple chances at breaking into
the torus. However, owing to the high optical depth of the torus, the
probability per attempt that \ac{IR} radiation can cross the entire torus is
very small, so most of the \ac{IR} radiation eventually leaves in the vertical
direction after a few ricochets off the inner surface. Through this process,
\ac{IR} radiation transfers its momentum several times to a thin layer of gas
at the inner surface.

This focusing of \ac{IR} radiative flux into the vertical direction means
$F_\su{IR}/F_\su{UV}$ rises gradually with $r$ in the central hole, as seen in
the bottom panel of \cref{fig:overview}. A consequence is that although the
wind is launched by \ac{UV} radiation, \ac{IR} radiation also contributes to
its acceleration once it reaches altitudes comparable to the vertical extent of
the torus.

We investigate $\vec F_\su{IR}$ more quantitatively with \cref{fig:IR radiative
flux}. The top panel displays $[L_\su{UV}/(4\pi r^2)]^{-1}\,\uvec e_r\cdot\vec
F_\su{IR}$ for $L_\su{UV}/L_\su E=0.11$ along lines emanating from the central
source at various inclinations; this quantity would be unity if the \ac{IR}
radiative flux were spherically symmetric. The solid portions of the curves
highlight the parts of the lines belonging to the torus proper. The lines are
divided into two classes. Lines at high inclinations pass through the torus and
have flux magnitudes below the spherically symmetric value. Conversely, lines
at low inclinations lie completely within the central hole and have flux
magnitudes above the spherically symmetric value; in fact, the curves appear to
converge to $\mathrelp\sim C_\su{IR}/(1-C_\su{IR})$ at large $r$, where
$C_\su{IR}$ is the \ac{IR} covering fraction (\cref{sec:peak IR energy
estimate}). The increasing discrepancy from spherical symmetry toward the
mid-plane illustrates the high degree of flux anisotropy.

\begin{figure}[!t]
\includegraphics{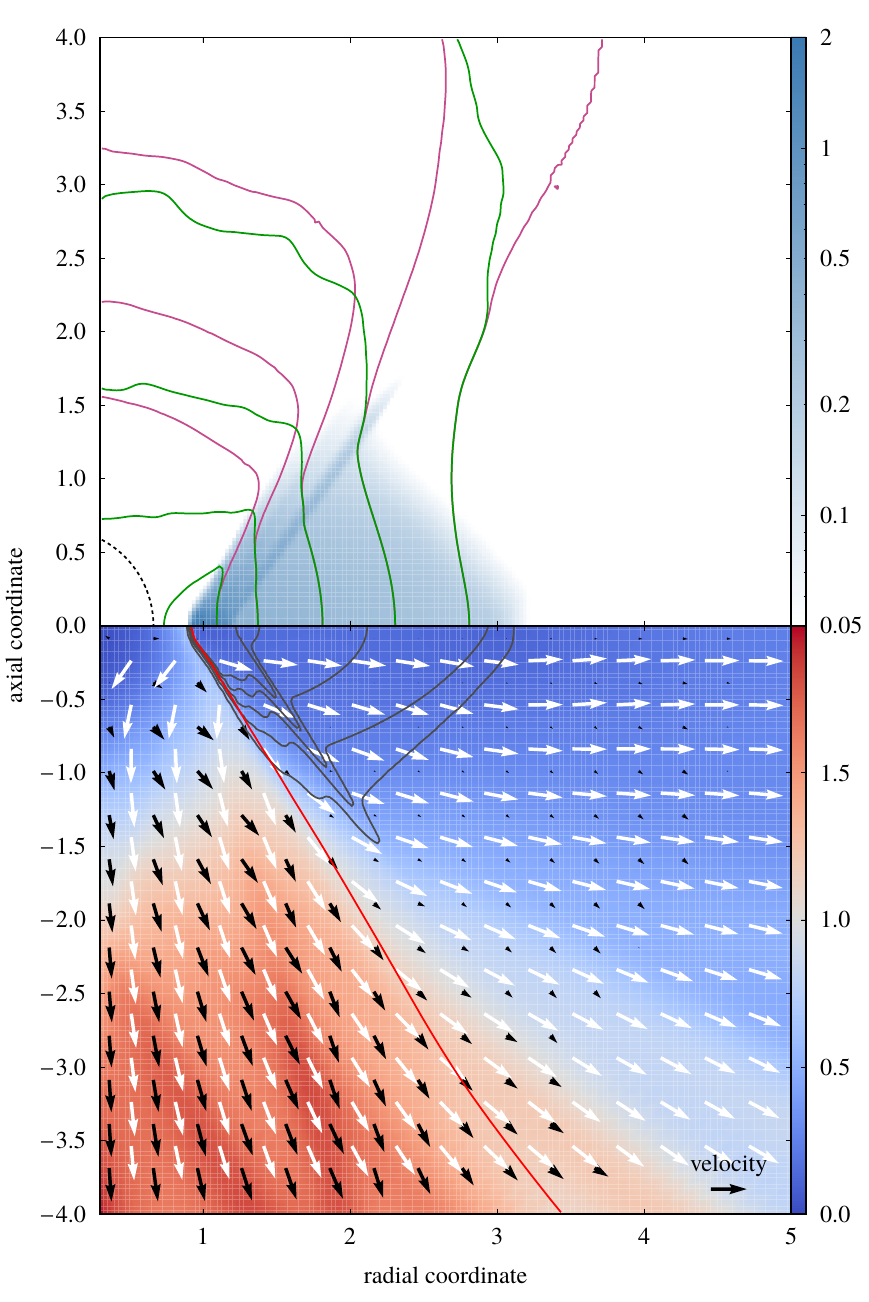}
\caption{Azimuthally averaged poloidal plane of the $L_\su{UV}/L_\su E=0.11$
simulation at time $t=2\,t_0$, but extending farther than in the top grid of
\cref{fig:time sequence}. All quantities are normalized to fiducial units
(\cref{sec:normalization}). \textit{Top panel:} Gas density is presented on a
logarithmic scale as blue intensities (see color bar along the right edge). The
dust sublimation surface $r=r_\su{ds}$ (\cref{sec:normalization}) is the dashed
black curve around the origin. Purple and green contours respectively show gas
and \ac{IR} radiation temperatures, both going from $0.3\,T_\su{ds}$ to
$0.7\,T_\su{ds}$ in steps of $0.1\,T_\su{ds}$ as one moves from the right to
the left; to avoid confusion, contours not passing through the torus body are
hidden. \textit{Bottom panel:} The background colors display $[L_\su{UV}/(4\pi
r^2)]^{-1}|F_\su{IR}$ (see color bar along the right edge), which is unity for
spherically symmetric radiation. The gray contours plot density rising from
$0.1\,\rho_0$ on the outside to $0.5\,\rho_0$ on the inside in steps of
$0.1\,\rho_0$. The $\tau_\su{UV}=1$ surface is traced by a red contour. The
white and black arrows graph $\vec F_\su{IR}/F_\su{IR}$ and $\vec v$
respectively; the arrow in the bottom-right corner has length $5\,v_0$.}
\label{fig:overview}
\end{figure}

\begin{figure}
\includegraphics{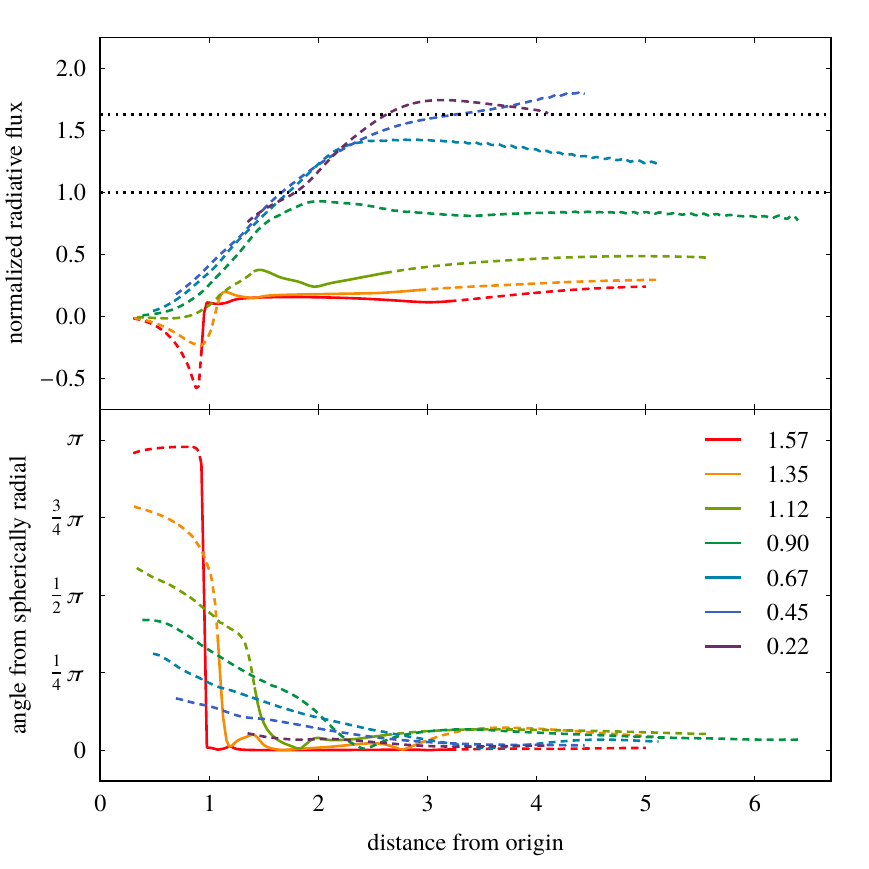}
\caption{Azimuthally averaged \ac{IR} radiative flux in the $L_\su{UV}/L_\su
E=0.11$ simulation at time $t=2\,t_0$ measured along lines with inclinations
indicated in the legend. The solid portion of each curve terminates at
$\tau_\su{UV}=1$ on the left and $\rho=(\bar\kappa_\su{IR}|r_0)^{-1}$ on the
right. All quantities are normalized to fiducial units
(\cref{sec:normalization}). \textit{Top panel:} Plot of $[L_\su{UV}/(4\pi
r^2)]^{-1}\,\uvec e_r\cdot\vec F_\su{IR}$; the upper and lower horizontal
dotted lines are drawn at $C_\su{IR}/(1-C_\su{IR})$ and 1. \textit{Bottom
panel:} Plot of $\arccos(\uvec e_r\cdot\vec F_\su{IR}/F_\su{IR})$.}
\label{fig:IR radiative flux}
\end{figure}

The bottom panel of \cref{fig:IR radiative flux} shows the angle between $\vec
F_\su{IR}$ and $\uvec e_r$. For lines at low inclinations, $\vec F_\su{IR}$ is
roughly parallel to $\uvec e_r$ everywhere; for lines at high inclinations, it
is intriguing that the \ac{IR} radiative flux snaps immediately to $\uvec e_r$
past the $\tau_\su{UV}=1$ surface. The fact that $\vec F_\su{IR}$ is nearly
aligned with $\uvec e_r$ in the body is all the more striking considering that
the \ac{IR} optical depth from the inner edge to the outer surface in the
vertical direction is a quarter that in the radial direction
(\cref{sec:normalization}).

Similar conclusions were also reached by \citet{2012ApJ...759...36R}, who found
that, for a smooth torus with geometrical thickness under a certain threshold,
most of the bolometric radiative flux exits through the central hole while only
a small fraction traverses the body. In addition, because $\vec
F_\su{UV}\propto\uvec e_r$ by definition, the bolometric radiative flux is
likewise spherically radial except where $\vec F_\su{IR}$ deviates most from
spherically radial, that is, just inside the uppermost parts of the inner
surface.

\Citet{2012ApJ...759...36R} also stated that $F_\su{IR}\propto r^{-2}$ at large
$r$. The top panel of \cref{fig:IR radiative flux} certainly suggests such a
trend, especially for \ac{IR} radiation beyond the outer surface. Nevertheless,
since the radial coordinate ratio of the outer to inner edge is small, we
cannot say with confidence if the inverse-square law holds inside the body. The
situation is also complicated by the torus not being in a quasi-steady state.

\subsection{Mass, momentum, and kinetic energy loss rates}
\label{sec:loss rates}

It is natural to ask how much mass, momentum, and kinetic energy are carried
away by the \ac{UV}\nobreakdash-launched wind mentioned in \cref{sec:gas
overview}. The rate at which mass is evacuated allows us to determine the
ultimate fate of the torus by balancing it against possible mass resupply.
Moreover, we can connect the loss rates in our simulations with observations of
\ac{AGN} outflows.

We emphasize that the chevron-shaped transient (\cref{sec:gas overview}) is not
the wind, and that the density concentration along the inner surface
(\cref{sec:gas overview}) does not trace the trajectory of individual gas
packets. Since the wind encompasses a large solid angle and density range, we
have no reliable way of separating it from the torus body, which is moving
radially outward at the same time along the mid-plane. In practice, we define
the mass loss rate as
\ifaastex{
  \begin{equation}
  \dot M\eqdef\int_{R=R_\su{max},\,\abs z>r_0}
    R\,d\phi\,dz\,\uvec e_R\cdot(\rho|\vec v)
    -\int_{z=z_\su{min}}R\,dR\,d\phi\,\uvec e_z\cdot(\rho|\vec v)
    +\int_{z=z_\su{max}}R\,dR\,d\phi\,\uvec e_z\cdot(\rho|\vec v),
  \end{equation}
}{
  \begin{align}
  \nonumber
  \dot M &\eqdef \int_{R=R_\su{max},\,\abs z>r_0}
    R\,d\phi\,dz\,\uvec e_R\cdot(\rho|\vec v) \\
  \nonumber
    &\noeq -\int_{z=z_\su{min}}R\,dR\,d\phi\,\uvec e_z\cdot(\rho|\vec v) \\
    &\noeq +\int_{z=z_\su{max}}R\,dR\,d\phi\,\uvec e_z\cdot(\rho|\vec v),
  \end{align}
}
and the momentum and kinetic energy loss rates in a similar fashion; here
$R_\su{max}$ and $z_{\su{min},\su{max}}$ denote the coordinates of the
boundaries of the simulation domain. We must be mindful to terminate our
analysis before the \ac{IR} half--opening angle becomes too large and the wind
drops below $\abs z=r_0$ at the outer-radial boundary, at $t\gtrsim6\,t_0$. All
loss rates derived from the simulations are implicitly quadrupled to account
for our limited azimuthal extent (\cref{sec:normalization}).

We begin with an analytic estimate of the mass loss rate. Supposing that the
wind is propelled by \ac{UV} momentum and reaches into $\tau_\su{UV}\sim1$, the
mass loss rate may be estimated by either $\dot M\sim L_\su{UV}/(c|v_\infty)$
or
\begin{equation}
\dot M\sim2|\left(\frac{2\pi|R_\su{in}}{\rho_\su{in}|\kappa_\su{UV}}\right)|
  (\rho_\su{in}|v_\infty)=4\pi|\frac{R_\su{in}|v_\infty}{\kappa_\su{UV}}.
\end{equation}
These two estimates agree if the wind terminal speed is
\begin{equation}\label{eq:escape velocity}
v_\infty\eqdef\biggl(\frac{GM}{R_\su{in}}|\frac{L_\su{UV}}{L_\su E}|
  \frac{\kappa_\su{UV}}\kappaT\biggr)^{1/2}.
\end{equation}
It follows that
\begin{equation}\label{eq:mass loss rate}
\dot M\sim4\pi|
  \biggl(\frac{GM|R_\su{in}}{\kappaT^2}|\frac{L_\su{UV}}{L_\su E}\biggr)^{1/2}|
  \biggl(\frac{\kappa_\su{UV}}\kappaT\biggr)^{-1/2}
\end{equation}
and
\begin{equation}\label{eq:kinetic luminosity}
\frac{\dot M|v_\infty^2}{L_\su{UV}}=\frac{v_\infty}c=
  \biggl(\frac{GM}{c^2|R_\su{in}}|
  \frac{L_\su{UV}}{L_\su E}|\frac{\kappa_\su{UV}}\kappaT\biggr)^{1/2}.
\end{equation}
When appropriately rewritten, \cref{eq:escape velocity,eq:mass loss rate} will
also serve as the basis of our scaling relations for extrapolating our
simulation results to more astrophysically relevant values of $M$ and
$\kappa_\su{UV}$ (\cref{sec:realistic parameters}).

The top panel of \cref{fig:loss rate} demonstrates that, in keeping with this
simple picture, the mass loss rates in our simulations normalized by
$L_\su{UV}/(c|v_\infty)$ are of order unity and nearly the same for all
$L_\su{UV}/L_\su E$ until $t\sim 4\,t_0$.

\begin{figure}
\includegraphics{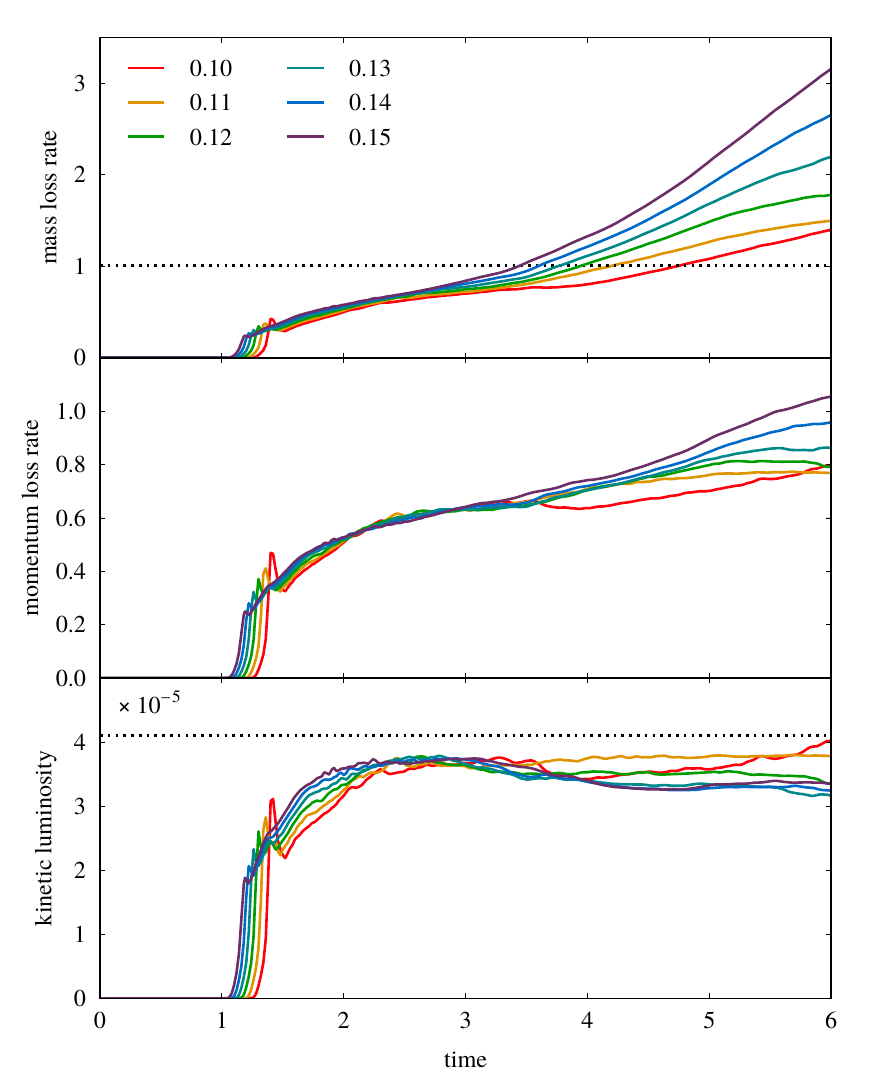}
\caption{\textit{Top panel:} Plot of the mass loss rate divided by
$L_\su{UV}/(c|v_\infty)$, with $v_\infty$ from \cref{eq:escape velocity}, for
each value of $L_\su{UV}/L_\su E$ indicated in the legend. The dotted line
shows the mass loss rate required to deplete an isolated torus within five
orbits if $L_\su{UV}/L_\su E=0.11$. \textit{Middle panel:} Plot of the
spherically radial gas momentum loss rate divided by $L_\su{UV}/c$.
\textit{Bottom panel:} Plot of the ratio of kinetic to \ac{UV} luminosity. The
dotted line shows the value of
$\ifaastex{v_\infty/(3c)}{\tfrac13|(v_\infty/c)}$ for $L_\su{UV}/L_\su E=0.11$;
the factor $\ifaastex{1/3}{\tfrac13}$ merely brings the line into the plot
range and has no physical meaning. All quantities are normalized to fiducial
units (\cref{sec:normalization}).}
\label{fig:loss rate}
\end{figure}

The middle panel traces the rate at which $\uvec e_r\cdot(\rho|\vec v)$, the
spherically radial component of gas momentum, leaves the simulation domain; the
normalization is $L_\su{UV}/c$, the rate of momentum injection in the form of
\ac{UV} radiation. This quantity is about half for $t_0\lesssim
t\lesssim4\,t_0$, suggesting that a sizable fraction of the radiation momentum
is not transferred to the gas.

We show in the bottom panel the ratio of kinetic to \ac{UV} luminosity, where
the kinetic luminosity is the loss rate of kinetic energy. Because
$R_\su{in}\propto M^{1/2}|(L_\su{UV}/L_\su E)^{1/2}$
(\cref{sec:normalization}), \cref{eq:kinetic luminosity} predicts $\dot
M|v_\infty^2/L_\su{UV}\propto M^{1/4}|(L_\su{UV}/L_\su E)^{1/4}$. The
$L_\su{UV}/L_\su E$ scaling is undetectable in our results since our range of
$L_\su{UV}/L_\su E$ spans a mere factor of 1.5; in fact, our ratio of kinetic
to \ac{UV} luminosity is effectively constant for all $L_\su{UV}$ simulated,
contrary to the $\mathrelp\propto L_\su{UV}^{1.8}$ scaling offered by
\citet{2012ApJ...759...36R}. Moreover, our explicit value is \num{\sim4e-3}
times that of \citet{2012ApJ...759...36R}, but this is largely because our $M$
is \num{\sim e-8} theirs and $\dot M|v_\infty^2/L_\su{UV}\propto M^{1/4}$.

\section{Discussion}
\label{sec:discussion}

We now interpret our simulation results and generalize them to
radiation-supported tori with different parameters.

\subsection{Estimation of \texorpdfstring{\acs*{IR}}{IR} radiation energy
density at the torus inner edge}
\label{sec:peak IR energy estimate}

The maximum of $E_\su{IR}$ is attained at the inner edge because that is where
\ac{UV} radiation is reprocessed (\cref{sec:IR overview}). We can estimate the
magnitude of the peak $(E_\su{IR})_\su{in\vphantom0}$ by considering the
radiation energy balance at the inner edge:
\begin{equation}\label{eq:inner edge energy balance}
C_\su{UV}|\frac{L_\su{UV}}{4\pi|R_\su{in}^2}
  +C_\su{IR}|(F_\su{IR}^-)_\su{in\vphantom0}\nosup
  \approx (F_\su{IR}^+)_\su{in\vphantom0}\nosup.
\end{equation}
We denote by $C_\su{IR,UV}$ the \ac{IR} and \ac{UV} covering fractions. Similar
to the two-stream approximation, we divide the radial component of the \ac{IR}
radiative flux into outward and inward parts, and we assign them to
$F_\su{IR}^\pm$ respectively. The second term on the left-hand side represents
the part of the \ac{IR} radiative flux leaking from the torus through the inner
edge into the central hole, and then absorbed at the far side after crossing
the hole.

\Cref{eq:inner edge energy balance} relates five variables at fixed $L_\su{UV}$
and is therefore difficult to verify against our simulations; two assumptions
simplify it. The first one is $C_\su{UV}\approx C_\su{IR}$. The second one
comes from observing that, for \ac{IR} optical depth $\Delta\tau_\su{IR}\gg1$
and covering fraction $C_\su{IR}\lesssim1$, \ac{IR} radiation propagates
diffusively at $R>R_\su{in}$, that is, $F_\su{IR}^++F_\su{IR}^-\approx
c|E_\su{IR}\gg F_\su{IR}^+-F_\su{IR}^-$, or
$F_\su{IR}^\pm\approx\ifaastex{c|E_\su{IR}/2}{\tfrac12|c|E_\su{IR}}$; we
suppose this holds at $R=R_\su{in}$ as well. \Cref{eq:inner edge energy
balance} then turns into
\begin{equation}\label{eq:peak IR energy estimate}
(E_\su{IR})_\su{in\vphantom0}\approx
  \frac{L_\su{UV}}{4\pi|R_\su{in}^2|c}|\frac{2|C_\su{IR}}{1-C_\su{IR}}.
\end{equation}
Our assumptions are not strictly correct because $C_\su{UV}>C_\su{IR}$, but
their errors act in opposite directions in such a way that \cref{eq:peak IR
energy estimate} is still an excellent description of our simulations. The
factor $C_\su{IR}/(1-C_\su{IR})$ is the number of scatterings \ac{IR} radiation
suffers at the inner surface prior to exit; $(E_\su{IR})_\su{in\vphantom0}$
goes up with $C_\su{IR}$ because the torus traps \ac{IR} more efficiently as
$C_\su{IR}$ approaches unity.

\Cref{fig:peak IR energy estimate} is a verification that, despite numerous
simplifications, \cref{eq:peak IR energy estimate} captures the physics well.
In preparation of this figure, we construct the radial profile of $E_\su{IR}$
by azimuthally averaging its mid-plane value; we then assign
$(E_\su{IR})_\su{in\vphantom0}$ and $R_\su{in}$ to the peak of the radial
profile and its radial coordinate respectively. We measure $C_\su{IR}$ by
considering a tight cylindrical envelope of the simulation domain, and
measuring the solid angle subtended at the origin by the parts of this envelope
for which the \ac{IR} optical depth toward the origin is greater than unity.
The success of \cref{eq:peak IR energy estimate} confirms the inner surface
does act like a mirror to \ac{IR} radiation.

\begin{figure}
\includegraphics{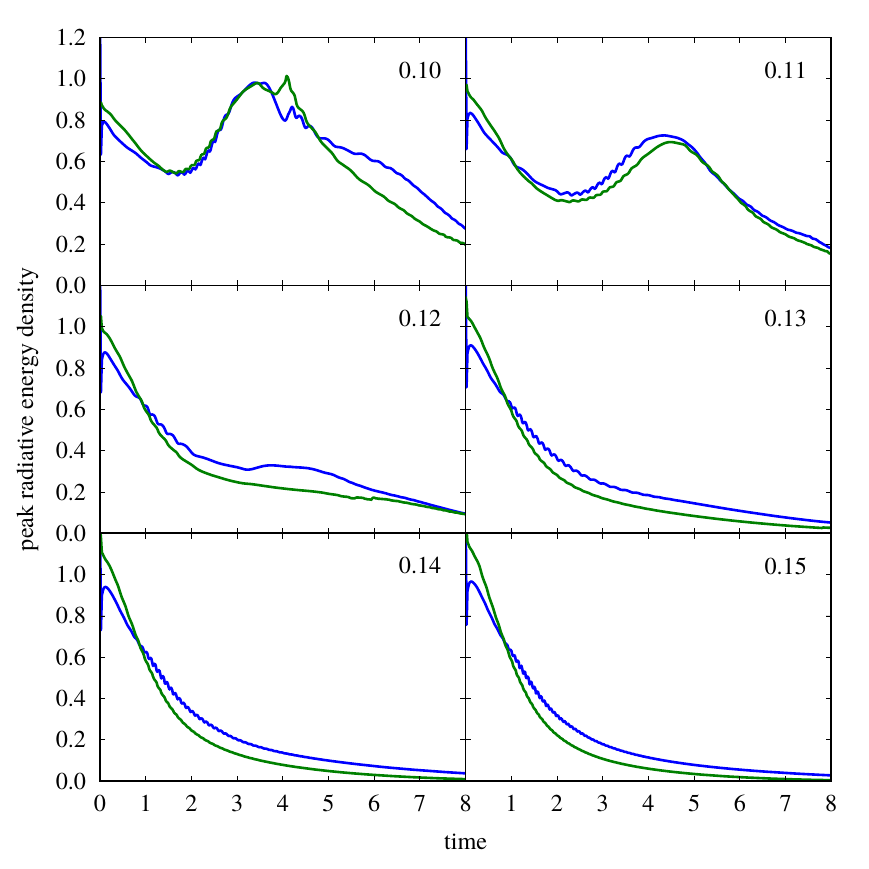}
\caption{Plot of the left- and right-hand sides of \cref{eq:peak IR energy
estimate} in blue and green curves respectively for each value of
$L_\su{UV}/L_\su E$ indicated in the top-right corner of each panel. All
quantities are normalized to fiducial units (\cref{sec:normalization}).}
\label{fig:peak IR energy estimate}
\end{figure}

Our study of the \ac{IR} covering fraction leads to another useful result: We
can predict the value of $L_\su{UV}$ that marginally balances gravity in our
initial condition (\cref{sec:initial condition}). Using \cref{eq:IC inner edge
energy ratio,eq:peak IR energy estimate}, we get
\begin{equation}
\frac{L_\su{UV}}{L_\su E}\approx
  \frac{\rho_\su{in}}{\rho_0}|\frac{R_\su{in}}{r_0}|
  \frac{1-C_\su{IR}}{2|C_\su{IR}}.
\end{equation}
The \ac{IR} half--opening angle at $t=0$ is \SI{\approx0.727}{\radian}; the
estimate $L_\su{UV}/L_\su E\approx0.135$ agrees with what we have found in
\cref{sec:gas overview}.

\subsection{Variation of simulation parameters}

It is useful to extend beyond the tiny parameter space explored by our
simulations. At fixed $M$, the principal parameters of the system are
$L_\su{UV}/L_\su E$, as well as $\rho_\su{in}$ and $j_\su{in}$ in the initial
condition (\cref{sec:initial condition}). We ignore detailed mass and angular
momentum distributions, although interesting local effects may arise if we
consider them fully. We also suppose the inner edge has temperatures near dust
sublimation \citep[e.g.][]{1969Natur.223..788R, 1981ApJ...250...87R,
1987ApJ...320..537B, 1989ApJ...337..236C, 1989ApJ...347...29S,
1993ApJ...418..673P}, so $R_\su{in}$ is not a free parameter once $M$ and
$L_\su{UV}/L_\su E$ are given. All these parameters enter into the net
acceleration
\begin{equation}
\vec a\eqdef-\frac{GM}{r^2}\,\uvec e_r+j^2|\frac{GM}{R^2}\,\uvec e_R+
  \frac{\kappa_\su{IR}}c|\vec F_\su{IR}+\frac{\kappa_\su{UV}}c|\vec F_\su{UV},
\end{equation}
which is a crucial factor governing torus dynamics. As far as global dynamics
are concerned, it is essentially a poloidal vector with radial and vertical
components $a_R\eqdef\uvec e_R\cdot\vec a$ and $a_z\eqdef(\sign z)\,\uvec
e_z\cdot\vec a$.

Consider how each parameter affects $a_R$ and $a_z$ in the torus body. Clearly
$a_R$ increases with $j_\su{in}$, while $a_R$ and $a_z$ increase with
$L_\su{UV}/L_\su E$ through $\vec F_\su{IR}$ and $\vec F_\su{UV}$. The
influence of $\rho_\su{in}$ on $a_R$ and $a_z$ is subtler as it simultaneously
controls $\Delta\tau_\su{IR}$ and $C_\su{IR}$, which play a role when
$\Delta\tau_\su{IR}\gtrsim1$. On the one hand, greater $\Delta\tau_\su{IR}$
reduces $F_\su{IR}$ in the body according to $F_\su{IR}\sim
c|(E_\su{IR})_\su{in\vphantom0}/\Delta\tau_\su{IR}$; on the other hand, greater
$C_\su{IR}$ better traps \ac{IR} radiation within the central hole, which at
constant $L_\su{UV}/L_\su E$ raises $(E_\su{IR})_\su{in\vphantom0}$
(\cref{sec:peak IR energy estimate}) and thus $F_\su{IR}$ in the body. Both
$\Delta\tau_\su{IR}$ and $C_\su{IR}$ rise with $\rho_\su{in}$, so it is
difficult to determine which effect dominates. In short, raising $j_\su{in}$
increases radial support, raising $L_\su{UV}/L_\su E$ increases both radial and
vertical support, whereas raising $\rho_\su{in}$ has an indeterminate effect on
support.

We now turn to local effects that can appear at the inner surface. First,
consider two tori with different $L_\su{UV}/L_\su E$ and $\rho_\su{in}$ tuned
so that they share $\vec F_\su{IR}$ and $\vec a$ in the body. Dynamics in the
body may be identical, but $\vec F_\su{IR}$ in the central hole of the torus
with greater $L_\su{UV}/L_\su E$ is necessarily stronger. Since \cref{eq:escape
velocity,eq:mass loss rate} show that $v_\infty$ and $\dot M$ depend on
$L_\su{UV}/L_\su E$ but not $\rho_\su{in}$, this torus must host a faster wind
than the other, as well as more severe losses of mass, momentum, and kinetic
energy. Second, a sufficiently large increase in either $L_\su{UV}/L_\su E$ or
$C_\su{IR}$ could make $(E_\su{IR})_\su{in\vphantom0}\gtrsim\aSB|T_\su{ds}^4$
and $\kappa_\su{IR}\approx0$ at the inner edge.

\subsection{Scaling simulation results to more realistic parameters}
\label{sec:realistic parameters}

As already remarked, for numerical reasons we have adopted artificially reduced
values of $M$ and $\kappa_\su{UV}/\kappa_\su{IR}$. Our tiny $M$ is the result
of requiring $c_\su s/v_\phi$ to be small, but not nearly as small as it would
be in a real system (\cref{sec:central mass}). The thickness of the \ac{UV}
radiation absorption layer at the inner surface is
$\mathrelp\lesssim(\kappa_\su{IR}/\kappa_\su{UV})|\Delta\tau_\su{IR}^{-1}$
times the radial extent of the torus, so a large opacity ratio would entail the
use of a grid size small enough to resolve an extremely thin absorption layer.
Moreover, because all of the momentum in \ac{UV} radiation is delivered within
the layer, gas in the layer experiences an acceleration
$\mathrelp\propto\kappa_\su{UV}$. With greater $\kappa_\su{UV}$, tracking the
development of the inner surface would necessitate high temporal resolution,
and the value of $\hat c$ would also need to be revised upward to keep $v<\hat
c$.

It is of course desirable to explore how the properties of our simulations
might change if those two parameters were pushed to astrophysically realistic
values. The true opacity ratio should be \numrange{\sim e2}{\sim e3}
\citep[e.g.,][]{2003A&A...410..611S}, but since the essential requirement for
capturing the physics is that the correct ordering of $\kappa_\su{IR}$ and
$\kappa_\su{UV}$ be kept, we argue our simulations are undamaged by our reduced
opacity ratio. To explore the effect of altering $\bar\kappa_\su{UV}$, we have
experimented with two simulations at twice the normal spatial resolution, one
with the usual value of $\bar\kappa_\su{UV}$, the other with twice the value.
The inner surface recedes slightly faster and is sharper at higher
$\bar\kappa_\su{UV}$, but otherwise the overall evolution of the torus and its
qualitative features are unaffected.

Nevertheless, quantitative results do vary with $\kappa_\su{UV}$; in
particular, care must be taken when scaling the wind terminal speed and mass
loss rate found in \cref{sec:loss rates}. A higher value of $\kappa_\su{UV}$
means the optically thin wind is faster but restricted to a thinner layer.
Rewriting \cref{eq:escape velocity,eq:mass loss rate} in terms of
$R_\su{in}/r_\su{ds}$ highlights how this scaling should be performed:
\ifaastex{
  \begin{equation}
  v_\infty\sim(GM|\kappaT|\aSB|T_\su{ds}^4)^{1/4}|
    \biggl(\frac{L_\su{UV}}{L_\su E}\biggr)^{1/4}|
    \biggl(\frac{\kappa_\su{IR}|\kappa_\su{UV}}{\kappaT^2}\biggr)^{1/4}|
    \biggl(\frac{R_\su{in}}{r_\su{ds}}\biggr)^{-1/2}
  \end{equation}
}{
  \begin{multline}
  v_\infty\sim(GM|\kappaT|\aSB|T_\su{ds}^4)^{1/4}\times{} \\
  \biggl(\frac{L_\su{UV}}{L_\su E}\biggr)^{1/4}|
    \biggl(\frac{\kappa_\su{IR}|\kappa_\su{UV}}{\kappaT^2}\biggr)^{1/4}|
    \biggl(\frac{R_\su{in}}{r_\su{ds}}\biggr)^{-1/2}
  \end{multline}
}
and
\ifaastex{
  \begin{equation}
  \dot M\sim4\pi|\left[\frac{(GM)^3}{\kappaT^5|\aSB|T_\su{ds}^4}\right]^{1/4}|
    \biggl(\frac{L_\su{UV}}{L_\su E}\biggr)^{3/4}|
    \biggl(\frac{\kappa_\su{IR}|\kappa_\su{UV}}{\kappaT^2}\biggr)^{-1/4}|
    \biggl(\frac{R_\su{in}}{r_\su{ds}}\biggr)^{1/2}.
  \end{equation}
}{
  \begin{multline}
  \dot M\sim4\pi|\left[\frac{(GM)^3}{\kappaT^5|\aSB|T_\su{ds}^4}\right]^{1/4}
    \times{} \\
  \biggl(\frac{L_\su{UV}}{L_\su E}\biggr)^{3/4}|
    \biggl(\frac{\kappa_\su{IR}|\kappa_\su{UV}}{\kappaT^2}\biggr)^{-1/4}|
    \biggl(\frac{R_\su{in}}{r_\su{ds}}\biggr)^{1/2}.
  \end{multline}
}
These forms cleanly separate the dependence on $M$ and $\kappa_\su{UV}$ from
everything else.

Shifting the fiducial values of these parameters from those used in our
simulations to more astrophysical numbers changes the wind terminal speed and
mass loss rate found in our simulations to
\ifaastex{
  \begin{equation}
  v_\infty\sim\num{5000}\,
    \biggl(\frac M{\SI{e7}{\solarmass}}\biggr)^{1/4}|
    \biggl(\frac{L_\su{UV}/L_\su E}{0.1}\biggr)^{1/4}|
  \biggr(\frac{\kappa_\su{IR}/\kappaT}{20}\biggr)^{1/4}|
    \biggr(\frac{\kappa_\su{UV}/\kappaT}{2000}\biggr)^{1/4}|
    \biggl(\frac{R_\su{in}}{r_\su{ds}}\biggr)^{-1/2}\,
    \si{\kilo\meter\per\second}
  \end{equation}
}{
  \begin{multline}
  v_\infty\sim\num{5000}\,
    \biggl(\frac M{\SI{e7}{\solarmass}}\biggr)^{1/4}|
    \biggl(\frac{L_\su{UV}/L_\su E}{0.1}\biggr)^{1/4}\times{} \\
  \biggr(\frac{\kappa_\su{IR}/\kappaT}{20}\biggr)^{1/4}|
    \biggr(\frac{\kappa_\su{UV}/\kappaT}{2000}\biggr)^{1/4}|
    \biggl(\frac{R_\su{in}}{r_\su{ds}}\biggr)^{-1/2}\,
    \si{\kilo\meter\per\second}
  \end{multline}
}
and
\ifaastex{
  \begin{equation}\label{eq:explicit mass loss rate}
  \dot M\sim0.1\,
    \biggl(\frac M{\SI{e7}{\solarmass}}\biggr)^{3/4}|
    \biggl(\frac{L_\su{UV}/L_\su E}{0.1}\biggr)^{3/4}|
  \biggr(\frac{\kappa_\su{IR}/\kappaT}{20}\biggr)^{-1/4}|
    \biggr(\frac{\kappa_\su{UV}/\kappaT}{2000}\biggr)^{-1/4}|
    \biggl(\frac{R_\su{in}}{r_\su{ds}}\biggr)^{1/2}\,
    \si{\solarmass\per\year}.
  \end{equation}
}{
  \begin{multline}\label{eq:explicit mass loss rate}
  \dot M\sim0.1\,
    \biggl(\frac M{\SI{e7}{\solarmass}}\biggr)^{3/4}|
    \biggl(\frac{L_\su{UV}/L_\su E}{0.1}\biggr)^{3/4}\times{} \\
  \biggr(\frac{\kappa_\su{IR}/\kappaT}{20}\biggr)^{-1/4}|
    \biggr(\frac{\kappa_\su{UV}/\kappaT}{2000}\biggr)^{-1/4}|
    \biggl(\frac{R_\su{in}}{r_\su{ds}}\biggr)^{1/2}\,
    \si{\solarmass\per\year}.
  \end{multline}
}
Outflows with speeds from \SIrange{\sim100}{\sim2000}{\kilo\meter\per\second}
have been identified in observations of X\nobreakdash-ray warm absorbers
\citep[e.g.,][]{2000A&A...354L..83K, 2000ApJ...535L..17K} and \ac{UV} absorbers
\citep[e.g.,][]{1969ApJ...158..859A, 1999ApJ...516..750C} in Seyfert~1s. Mass
loss rates inferred from X\nobreakdash-ray warm absorbers go from
\SIrange{\sim e-3}{\sim10}{\solarmass\per\year}
\citep[e.g.,][]{2005A&A...431..111B, 2011MNRAS.410.2274Z}, whereas studies of
\ac{UV} absorbers suggest a wider range of \SIrange{\sim
e-4}{\sim10}{\solarmass\per\year} \citep{2012ApJ...753...75C}. These empirical
results are roughly consistent with our fiducial values of $v_\infty$ and $\dot
M$.

The mass loss rate can be understood in a more intuitive fashion. The initial
mass of the torus is $M_\su{tor}\eqdef
C_\su{tor}\times2\pi|\rho_\su{in}|R_\su{in}^3$, where $C_\su{tor}\approx4$ for
our initial condition (\cref{sec:initial condition}). We define the lifetime of
the torus against mass loss in the radiation-driven wind as $t_\su{tor}\eqdef
M_\su{tor}/\dot M$; from \cref{eq:mass loss rate}, we have
\begin{equation}\label{eq:torus lifetime}
t_\su{tor}|\Omega_\su{in}\sim\frac12|C_\su{tor}|
  (\tau_\su T\nosup|\tau_\su{UV}\nosup)_\su{in\vphantom0}^{1/2}|
  \left(\frac{L_\su{UV}}{L_\su E}\right)^{-1/2}.
\end{equation}
In this equation, $\Omega_\su{in}=(GM/R_\su{in}^3)^{1/2}$ is the orbital
frequency at the inner edge, and $(\tau_\su{T,UV})_\su{in\vphantom0}\eqdef
\rho_\su{in}|R_\su{in}|\kappa_\su{T,UV}$ stand for Thomson and \ac{UV} optical
depths evaluated with inner-edge values. Our simulations have $(\tau_\su
T)_\su{in\vphantom0}\sim0.8$, $(\tau_\su{UV})_\su{in\vphantom0}\sim64$, and
$0.10\le L_\su{UV}/L_\su E\le0.15$, so $t_\su{tor}|\Omega_\su{in}\sim42$. Our
torus remains inside the simulation domain for a shorter amount of time because
the torus body moves radially outward at late times (\cref{sec:gas overview}).
Our Thomson optical depth may be reasonable for real \acp{AGN}, but our \ac{UV}
optical depth is too small by a factor of \num{\gtrsim10}, so we expect the
lifetime of realistic tori against mass loss to be \num{\gtrsim3} times longer.

\subsection{Balance between radiation-driven mass loss and mass resupply}
\label{sec:resupply}

A salient feature of simulations for all $L_\su{UV}$ is a radiation-driven wind
from the inner surface (\cref{sec:gas overview}). The wind always has
temperatures below $T_\su{ds}$ since it lies outside of the dust sublimation
radius. Depending on the geometry of the inner edge, the wind can be found at
higher latitudes than the torus body; this enhances the covering fraction, and
hints at a connection between the wind and dust observed in the polar regions
of \ac{NGC}~424 \citep{2012ApJ...755..149H} and \ac{NGC}~3783
\citep{2013ApJ...771...87H}.

The radiation-driven wind is distinct from the thermally driven wind
\citep{1983ApJ...271...70B} commonly discussed in the context of the torus
\citep{1986ApJ...308L..55K, 2001ApJ...561..684K, 2005A&A...431..111B}. The
latter refers to gas lifted from the inner surface, exposed to ionizing
radiation from the central source, and heated to the Compton temperature soon
after its ionization parameter exceeds unity \citep{1981ApJ...249..422K}. The
mass loss rate due to the thermally driven wind is
\SI{\sim0.4}{\solarmass\per\year} \citep{1986ApJ...308L..55K}, similar to that
of the radiation-driven wind found in \cref{eq:explicit mass loss rate}. The
two winds could consequently augment each other despite their different
physical properties.

The mass lost to these winds could be resupplied from the outside. A steady
state could also obtain in which the \ac{IR} optical depth across the body is
approximately constant, so that the \ac{IR} radiative flux does not become
powerful enough to shove the body collectively outward (\cref{sec:gas
overview}). A combined molecular and ionized gas inflow rate of
\SI{\sim0.2}{\solarmass\per\year} has been observed down to
\SI{\sim40}{\parsec} in \ac{NGC}~1097 \citep{2013ApJ...770L..27F}. Inflows of
this magnitude at the outskirts of the torus suffice to replenish the mass loss
given by \cref{eq:explicit mass loss rate}.

Magnetic effects can strongly influence the resupply rate. \Ac{MHD} turbulence
stirred up by the \ac{MRI} could lead to outward angular momentum transport
through the torus and subsequent accretion toward the inner edge. The ideal
\acp{MHD} condition holds even at extremely low ionization fractions
\citep{1994ApJ...421..163B, 1996ApJ...457..355G}, which can be maintained by
X\nobreakdash-rays \citep{1995ApJ...447L..17N} if they carry a sizeable
fraction of the energy in the \ac{UV} \citep[e.g.,][]{1981ApJ...245..357Z}.
Indeed, magnetic fields have been detected on \SI{\lesssim30}{\parsec} scales
in the nucleus of \ac{NGC}~1068 \citep{2015MNRAS.452.1902L}.

Recall that the steady-state mass inflow timescale in a disk is
$\mathrelp\sim[\alpha|(H/R)^2|\Omega]^{-1}$, where $H/R$ and $\Omega$ are the
aspect ratio and orbital frequency of the disk. Accretion driven by \ac{MHD}
stresses has $0.01\lesssim\alpha\lesssim0.1$, so the inflow timescale is quite
close to the torus lifetime calculated in \cref{eq:torus lifetime} if, as here,
$H/R\sim 1$. The relatively mild dependence of $t_\su{tor}|\Omega_\su{in}$ on
$L_\su{UV}/L_\su E$ suggests that equilibrium between inflow and outflow could
be attained over a wide range of luminosities.

The presence of \ac{MHD} stresses can redistribute angular momentum in the
torus, altering the distribution of \ac{IR} radiation needed to achieve radial
force balance against gravity; this change could in turn affect whether the
torus is vertically supported. Magnetic fields could also remove angular
momentum altogether from the torus through a magnetized wind
\citep{1982MNRAS.199..883B, 1994ApJ...434..446K}.

\subsection{Radial perturbation of the torus inner surface}
\label{sec:inner surface perturbation discussion}

The nonlinear development of the radial perturbation of the inner surface
(\cref{sec:inner surface perturbation}) is reminiscent of the Rayleigh--Taylor
instability \citep{1883PLMS...14..170R!, 1950RSPSA.201..192T}. For
$L_\su{UV}/L_\su E\ge0.12$, the emergence of fingers and arcs from an
originally smooth inner surface is a hallmark of the instability. For
$L_\su{UV}/L_\su E=0.11$, the azimuthal wavenumber of the most prominent mode
of the perturbation decreases as the development of the perturbation becomes
nonlinear; this mirrors the classical picture in which the fastest-growing mode
of the perturbation of the interface separating the two fluids shifts from high
wavenumbers in the linear regime to low wavenumbers in the nonlinear regime
\citep[e.g.][]{1957RSPSA.241..423G, 1959PhFl....2..656C}.

Since radiation and not a physical fluid is supporting the gas against gravity,
it is more accurate to compare our simulations with the radiative
Rayleigh--Taylor instability \citep{1977PhFl...20..364K, 1977ApJ...214...10M,
2011ApJ...730..116J, 2013ApJ...763..102J}. \Cref{fig:RTI} shows that in both
linear and nonlinear regimes, wherever a part of the $\tau_\su{UV}=1$ surface
is farther from the origin, the region immediately radially outward of it has
greater $R^2\,\uvec e_R\cdot\vec F_\su{IR}$ because the optical depth to the
outer surface is smaller; the perturbation grows as a consequence. This
mechanism is similar to what \citet{1977PhFl...20..364K} described.
Nonetheless, the cylindrical geometry of our simulations, as well as the
presence of an acceleration gradient and differential rotation, complicates
direct comparison with these previous analyses.

The amplification of the perturbation turns a smooth density distribution
inhomogeneous; this kind of fragmentation process could provide a physical
mechanism for the formation of dusty clumps often invoked to explain the
observed broad \SIrange{\sim1}{\sim100}{\micro\meter} bump in the \ac{SED} of
\acp{AGN} \citep{2002ApJ...570L...9N, 2008ApJ...685..160N}, the weak
\SI{9.7}{\micro\meter} silicate emission or absorption feature
\citep{2002ApJ...570L...9N, 2008ApJ...685..160N, 2006A&A...452..459H}, and the
gentle radial temperature profile of dust within the central parsec of Circinus
\citep{2007A&A...474..837T}.

Radiation-driven clump formation has already been reported in \ac{FLD}
simulations of super-Eddington outflows from axisymmetric accretion disks
\citep{2013PASJ...65...88T} and from two-dimensional planar atmospheres
\citep{2014PASJ...66...48T} where the dominant source of opacity is electron
scattering. Clumps in these simulations are irregular and typically one optical
depth across. Anisotropic structures are likewise observed in our
three-dimensional simulations employing genuine \ac{RT}, but they have multiple
characteristic length scales. Magnetic fields certainly exist in the torus
\citep{2015MNRAS.452.1902L} and could change how fragments are formed and
destroyed, but we must leave its study to future work.

\section{Conclusions}

We have conducted three-dimensional, time-dependent \acp{RHD} simulations of
\ac{AGN} tori in which gas and radiation are evolved simultaneously, and
\ac{IR} and \ac{UV} radiative fluxes are not approximated using arbitrary
closure prescriptions. The simulations reveal that a smooth, geometrically and
Compton thick torus is not very permeable to \ac{IR} radiation, whereas the
optically thin central hole allows \ac{IR} radiation to escape immediately;
therefore, the \ac{IR} radiative flux is much stronger through the central hole
than across the torus, and \ac{IR} radiative support inside the torus is weaker
than if the torus body were optically thin (\cref{sec:IR overview}). Meanwhile,
\ac{IR} radiation undergoing several reflections at the inner surface before
leaving the central hole enhances the \ac{IR} radiation energy density at the
inner edge (\cref{sec:peak IR energy estimate}) and reduces the luminosity
needed to achieve marginal \ac{IR} radiative support.

The inner surface experiences a spontaneous breaking of axisymmetry under
radiation and differential rotation; the consequent radial perturbation
amplifies rapidly with time (\cref{sec:inner surface perturbation}). The growth
of the perturbation conjures up the picture of the radiative Rayleigh--Taylor
instability, but with critical differences. The fragmentation of the inner
surface alludes to a physical mechanism for the creation of clumps; however,
the steady-state configuration of the fragments is not probed by our
simulations and is likely affected by magnetic fields (\cref{sec:inner surface
perturbation discussion}).

Most importantly, a dusty wind can be launched from the inner surface by
\ac{UV} radiation and propelled outward by a combination of \ac{IR} and \ac{UV}
radiation. The appearance of this wind is inevitable in a torus with vertical
density stratification (\cref{sec:gas overview}). High dust opacity in the
\ac{UV}, along with the concentration of \ac{IR} radiative flux into the
vertical direction (\cref{sec:IR overview}), means the wind likely experiences
an acceleration well above gravity. The radiation-driven wind carries momentum
comparable to that in \ac{UV} radiation (\cref{sec:loss rates}). It is also a
powerful mechanism of mass loss with the capacity to remove an isolated torus
within \num{\sim20} orbital periods at the inner edge (\cref{sec:realistic
parameters}).

Our study calls attention for the first time to the possibility that \ac{UV}
radiation pressure acting on dust can drive a wind with speed and mass loss
rate of the same order as values inferred from observations
(\cref{sec:realistic parameters}), and with mass loss rate similar to the
better-known thermally driven wind (\cref{sec:resupply}). In order to achieve
an approximate steady state against mass loss through both kinds of winds, any
such torus must be furnished with a new inventory of mass every \num{\sim20}
orbital periods. The strong variation of radiative support throughout the body
(\cref{sec:gas overview}), the existence of a radiation-driven wind
(\cref{sec:gas overview,sec:loss rates}), and the growth of perturbations along
the inner surface (\cref{sec:inner surface perturbation}), demonstrate that the
internal structures of tori are unlikely ever to achieve strict hydrostatic
equilibrium.

\vskip\bigskipamount\noindent
The authors thank Jim Stone, Yanfei Jiang, and Shane Davis for generously
allowing Athena and its time-dependent \ac{RT} module to be used for this
project and for providing technical support. This research was partially
supported by NASA/ATP grants NNX11AF49G and NNX14AB43G\@. The simulations were
performed on the Johns Hopkins Homewood High-Performance Cluster.

\begin{appendices}

\section{Time-independent long-characteristics UV RT}
\label{sec:UV radiative transfer}

We have developed a time-independent long-characteristics \ac{RT} module to
deal with \ac{UV} radiation from a point source at the origin in cylindrical
coordinates. Our ray-casting algorithm is similar to that of
\citet{1987AmanatidesWoo}. We construct a ray from the source to the center of
every cell in the simulation domain, extend it so that it reaches the far side
of the destination cell, and then chop it up into segments, one for each cell
the ray passes through. This ray-casting is done once, before the simulation
starts. Our adoption of cylindrical coordinates means that we only need to
solve the ray-casting problem in two dimensions. A subtlety of our algorithm is
that, whenever a ray passes very close to a cell corner, we allow the ray to
pass diagonally through it.

At the beginning of a time step, we compute the \ac{UV} radiation energy
density in the destination cell by
\begin{equation}
\frac{4\pi}c|J_\su{UV}\eqdef\frac{L_\su{UV}}{4\pi r^2c}|e^{-\tau_\su{UV}}|
  \frac{\exp(\tfrac12\tau^*_\su{UV})-\exp(-\tfrac12\tau^*_\su{UV})}
  {\tau^*_\su{UV}},
\end{equation}
where $L_\su{UV}$ is the luminosity of the source in the \ac{UV} and $\vec r$
is the displacement from the source to the destination cell. We determine the
\ac{UV} optical depth $\tau_\su{UV}$ by accumulating the products of the length
of each segment and $\rho|\kappa_\su{UV}$ averaged over the cell in which the
segment lies; note that we consider only half of the length of the last segment
in this exercise. The last factor in the equation comes from averaging
$J_\su{UV}$ over the entire last segment, which has \ac{UV} optical depth
$\tau^*_\su{UV}$; its inclusion improves the agreement of the \ac{UV} energy
and momentum absorption rate between runs at different resolutions,
particularly at locations where $\tau_\su{UV}\lesssim1$.

To arrive at the energy and momentum source terms of gas due to \ac{UV}
radiation, we remind ourselves of the \ac{RT} equation in the form derived by
\citet{1982JCoPh..46...97M}:
\begin{multline}\label{eq:MK1982 radiative transfer}
\frac1c|\pd{I_\su{UV}}t+\uvec n\cdot\grad I_\su{UV}=
  \Bigl(-1+\uvec n\cdot\frac{\vec v}c\Bigr)|
  \rho|(\kappa_\su{UV}+\sigma_\su{UV})|I_\su{UV} \\
+\Bigl(1+3\,\uvec n\cdot\frac{\vec v}c\Bigr)|
  \rho|(\kappa_\su{UV}|B+\sigma_\su{UV}|J_\su{UV})
  -2|\rho|\sigma_\su{UV}|\frac{\vec v}c\cdot\vec H_\su{UV}.
\end{multline}
The zeroth and first angular moments of \cref{eq:MK1982 radiative transfer} are
\ifaastex{
  \begin{alignat}{2}
  \label{eq:MK1982 radiative zeroth moment}
  \frac1c|\pd{J_\su{UV}}t+\divg\vec H_\su{UV} &=
    \rho|\kappa_\su{UV}|(B-J_\su{UV})+
    \rho|(\kappa_\su{UV}-\sigma_\su{UV})|\frac{\vec v}c\cdot\vec H_\su{UV}
    &&\eqdef -\frac1{4\pi}|\ergsrc{UV}, \\
  \label{eq:MK1982 radiative first moment}
  \frac1c|\pd{\vec H_\su{UV}}t+\divg\tsr K_\su{UV} &=
    -\rho|(\kappa_\su{UV}+\sigma_\su{UV})|
    \Bigl(\vec H_\su{UV}-\frac{\vec v}c\cdot\tsr K_\su{UV}\Bigr)
    +\frac{\vec v}c|\rho|(\kappa_\su{UV}|B+\sigma_\su{UV}|J_\su{UV})
    &&\eqdef -\frac c{4\pi}|\momsrc{UV}.
  \end{alignat}
}{
  \begin{alignat}{2}
  \nonumber
  & \frac1c|\pd{J_\su{UV}}t+\divg\vec H_\su{UV}= \\
  \label{eq:MK1982 radiative zeroth moment}
  &\quad \rho|\kappa_\su{UV}|(B-J_\su{UV})
    +\rho|(\kappa_\su{UV}-\sigma_\su{UV})|\frac{\vec v}c\cdot\vec H_\su{UV}
    &&\eqdef -\frac1{4\pi}|\ergsrc{UV}, \\
  \nonumber
  & \frac1c|\pd{\vec H_\su{UV}}t+\divg\tsr K_\su{UV}= \\
  \nonumber
  &\quad \mathbinu-\rho|(\kappa_\su{UV}+\sigma_\su{UV})|
    \Bigl(\vec H_\su{UV}-\frac{\vec v}c\cdot\tsr K_\su{UV}\Bigr) \\
  \label{eq:MK1982 radiative first moment}
  && \mathllap{
    {}+\frac{\vec v}c|\rho|(\kappa_\su{UV}|B+\sigma_\su{UV}|J_\su{UV})}
    &\eqdef -\frac c{4\pi}|\momsrc{UV}.
  \end{alignat}
}
\Citet{1999ApJ...521..432L} pointed out that \cref{eq:MK1982 radiative
transfer,eq:MK1982 radiative zeroth moment,eq:MK1982 radiative first moment} do
not give the correct equilibrium in moving fluids. To overcome this problem,
the time-dependent \ac{RT} module of Athena solves the modified
\cref{eq:radiative transfer}, but \cref{eq:radiative transfer,eq:MK1982
radiative transfer} are identical to first order in $v/c$ save for the
subscripts.

For time-independent \ac{RT}, which applies to the \ac{UV}, we drop the time
derivatives from \cref{eq:MK1982 radiative transfer,eq:MK1982 radiative zeroth
moment,eq:MK1982 radiative first moment}. In the special case of point-source
\ac{UV} radiation interacting with purely absorbing material that does not
re-radiate in the \ac{UV}, we set $\sigma_\su{UV}=0$, $B=0$, $\vec
H_\su{UV}=\uvec e_r\,J_\su{UV}$, and $\tsr K_\su{UV}=\uvec e_r\,\uvec
e_r\,J_\su{UV}$ in \cref{eq:MK1982 radiative zeroth moment,eq:MK1982 radiative
first moment}; the source terms we seek can be skimmed off as
\begin{align}
-\frac1{4\pi}|\ergsrc{UV} &\eqdef
  -\rho|\kappa_\su{UV}|J_\su{UV}|\Bigl(1-\uvec e_r\cdot\frac{\vec v}c\Bigr), \\
-\frac c{4\pi}|\momsrc{UV} &\eqdef
  -\rho|\kappa_\su{UV}|J_\su{UV}\,\uvec e_r\,
  \Bigl(1-\uvec e_r\cdot\frac{\vec v}c\Bigr).
\end{align}
Observe that a consistent solution cannot be reached with \cref{eq:radiative
zeroth moment,eq:radiative first moment}.

As implied by \cref{eq:gas momentum,eq:gas energy}, the source terms are added
directly to the gas at the beginning of the time step. The energy source term
is rather large compared to the other terms of \cref{eq:gas energy}, so the gas
is temporarily overheated. The \ac{IR} radiative sub-step is then carried out
as described by \citet{2014ApJS..213....7J}, during which the gas releases
almost all of the energy it gained from the \ac{UV} into the \ac{IR}. Although
the source terms are added using the explicit Euler method, the \ac{IR}
radiative sub-step proceeds by the implicit Euler method, hence a large energy
source term does not pose a problem.

Despite the sharp rise in gas temperature after the \ac{UV}
long-characteristics sub-step, we must not change the \ac{IR} and \ac{UV}
opacities until the \ac{IR} radiative sub-step is finished; otherwise, gas
exposed to \ac{UV} radiation would be absorbing \ac{UV} and emitting \ac{IR} at
two unrelated opacities, which would generate specious temperature fluctuations
with a period equal to two or three time steps around the true equilibrium
value.

\section{Reduced speed of light approximation}
\label{sec:reduced light speed}

The radiative timescale governing \cref{eq:radiative transfer,eq:radiative
zeroth moment,eq:radiative first moment} is shorter than the hydrodynamic
timescale of \cref{eq:gas mass,eq:gas momentum,eq:gas energy} by a factor of
$c/v\gg1$. Our primary concern is the hydrodynamic timescale, whereas the fast
variation of $I_\su{IR}$ relative to $\rho$, $\vec v$, and $p$ is uninteresting
since radiation merely equilibriates with the gas in between hydrodynamic time
steps. To avoid following the system on the radiative timescale, we adopt the
method of reduced speed of light \citep{2001NewA....6..437G,
2013ApJS..206...21S}.

The physical light speed $c$ attached to the time derivatives in
\cref{eq:radiative transfer,eq:radiative zeroth moment,eq:radiative first
moment} is substituted with the reduced light speed $\hat c$. This allows the
use of coarser temporal resolution since the rate of change of $I_\su{IR}$ in
\cref{eq:radiative transfer}, including thermalization by absorption,
isotropization by scattering, propagation in vacuum, and advection in optically
thick gas, is slowed down by a factor of $\hat c/c$. The source terms
$\ergsrc{IR}$ and $\momsrc{IR}$ are not altered, only the rate at which they
change $J_\su{IR}$ and $\vec H_\su{IR}$ in \cref{eq:radiative zeroth
moment,eq:radiative first moment}; in fact, they must not be touched in
\cref{eq:gas momentum,eq:gas energy} if we are to preserve gas dynamics.

Our approximation does not stop radiation from reaching equilibrium with the
gas inasmuch as $v<\hat c\ll c$. However, it is critical that we not replace
$c$ attached to $\vec v/c$ in \cref{eq:radiative transfer,eq:radiative zeroth
moment,eq:radiative first moment}, otherwise $\vec H_\su{IR}$ could be beamed
in the direction of $\vec v$ even when $v\lesssim\hat c\ll c$.

Because the rate of change of energy and momentum of gas is $c/\hat c$ times
that of radiation, \cref{eq:gas momentum,eq:gas energy,eq:radiative zeroth
moment,eq:radiative first moment} taken together do not conserve the physical
values of energy and momentum, but $E+4\pi|J_\su{IR}/\hat c$ and $\rho|\vec
v+4\pi|\vec H_\su{IR}/(c\hat c)$ instead. Granted that spurious transients may
manifest themselves on approach to energy and momentum equilibrium between gas
and radiation, we nevertheless expect time-averaged values of $\ergsrc{IR}$ and
$\momsrc{IR}$ to vanish once equilibrium prevails.

A technical point to bear in mind is that the time-dependent \ac{RT} module of
Athena evaluates $\ergsrc{IR}\,\Delta t$ and $\momsrc{IR}\,\Delta t$ not from
the right-hand sides of \cref{eq:radiative zeroth moment,eq:radiative first
moment}, but directly from $\Delta I_\su{IR}$ as computed by the \ac{IR}
radiative sub-step; the conversion from $\Delta I_\su{IR}$ to
$\ergsrc{IR}\,\Delta t$ and $\momsrc{IR}\,\Delta t$ therefore necessitates a
factor of $c/\hat c$.

\section{Improvement to treatment of scattering in Athena}
\label{sec:scattering solution}

We consider the treatment of scattering opacity by the time-dependent \ac{RT}
module of Athena. The notation follows \citet{2014ApJS..213....7J}, except that
here $c$ and $\hat c$ are the physical and reduced light speeds. We define
$\sigma_\su s$ as the scattering cross section per volume, $\Delta t$ as the
time step, and $\zeta\eqdef\tau^*_\su s|v/c$, where $\tau^*_\su
s\eqdef\sigma_\su s|\hat c|\Delta t$.

The module handles scattering by solving equations (29) and (30) listed in the
reference. We repeat the equations below, minus a couple typos:
\begin{equation}
\ifaastex{}{\arraycolsep.03em\medmuskip-.2\medmuskip}
\begin{pmatrix}
  a_1+b_1+c_1 & b_2+c_1 & b_3+c_1 & \cdots & b_N+c_1 \\
  b_1+c_2 & a_2+b_2+c_2 & b_3+c_2 & \cdots & b_N+c_2 \\
  b_1+c_3 & b_2+c_3 & a_3+b_3+c_3 & \cdots & b_N+c_3 \\
  \vdots & \vdots & \vdots & \ddots & \vdots \\
  b_1+c_N & b_2+c_N & b_3+c_N & \cdots & a_N+b_N+c_N
\end{pmatrix}
\begin{pmatrix} x_1 \\ x_2 \\ x_3 \\ \vdots \\ x_N \end{pmatrix}=
\begin{pmatrix} r_1 \\ r_2 \\ r_3 \\ \vdots \\ r_N \end{pmatrix},
\end{equation}
where
\begin{subequations}\begin{align}
a_l &\eqdef W_l^{-1}|[1+\tau^*_\su s|(1-\vec n_l\cdot\vec v/c)], \\
b_l &\eqdef \tau^*_\su s|
  [2|(\vec n_l\cdot\vec v/c)-(\vec n_l\cdot\vec v/c)^2-v^2/c^2], \\
c_l &\eqdef -\tau^*_\su s|[1+3|(\vec n_l\cdot\vec v/c)], \\
x_l &\eqdef W_l|I^{n+1}_l, \\
r_l &\eqdef I^n_l.
\end{align}\end{subequations}
In the process of computing the $LU$ decomposition of the matrix,
$c_l-c_N\sim\zeta$ appears multiple times in the denominator. If $\zeta\ll1$,
some elements of the resultant matrices are $\num{\sim1}$ while others are
$\mathrelp\sim\zeta^{-1}$; put differently, the matrices are ill-conditioned,
with the ratio of the greatest to smallest singular values of either matrix
being $\mathrelp\sim\zeta$. Because the solution is computed using
back-substitution as $x_i=U\inv_{ij}|L\inv_{jk}|r'_k$, it could be highly
inaccurate. The code already includes checks to avoid this kind of situation,
but it is easy to construct triples of $\tau^*_\su s$, $v$, and $c$ that bypass
them.

When $\zeta$ is below a certain threshold, it is preferable to regard the
scattering equation, written in the form
\begin{equation}
I^n_l=\sum_{m=1}^N(\lambda_{lm}+\epsilon_{lm})|I^{n+1}_m,
\end{equation}
where
\begin{subequations}\begin{align}
\lambda_{lm} &\eqdef (1+\tau^*_\su s)|\krons{lm}-\tau^*_\su s|W_m, \\
\epsilon_{lm} &\eqdef \tau^*_\su s|(a'_l|\krons{lm}+b'_m|W_m+c'_l|W_m), \\
a'_l &\eqdef -\vec n_l\cdot\vec v/c, \\
b'_l &\eqdef 2|(\vec n_l\cdot\vec v/c)-(\vec n_l\cdot\vec v/c)^2, \\
c'_l &\eqdef -[3|(\vec n_l\cdot\vec v/c)+v^2/c^2],
\end{align}\end{subequations}
and $\krons{lm}$ is the Kronecker delta, as a perturbative equation and solve
it iteratively. The procedure starts with $I^{n+1}_l\leftarrow I^n_l$; each
iterative step updates the solution as
\begin{equation}
I^{n+1}_l\leftarrow\sum_{m=1}^N\lambda\inv_{lm}
  \left[I^n_m-\sum_{p=1}^N\epsilon_{mp}|I^{n+1}_p\right].
\end{equation}
The special structure of the matrix allows the inner multiplication to be
accomplished with time expenditure $\bigO(N)$, while the multiplication of any
vector $v_m$ by the inverse matrix $\lambda\inv_{lm}$ is simply
\begin{equation}
\sum_{m=1}^N\lambda_{lm}^{-1}|v_m
  =\frac1{1+\tau^*_\su s}|\left(v_l+\tau^*_\su s\sum_{m=1}^NW_m|v_m\right)
\end{equation}
since $\mathop{\smash{\sum_{m=1}^N}}W_m=1$. The chief aim of this modification
is not to obtain a more accurate solution when $\zeta\sim\num{e-15}$; rather,
it prevents the numerical instability that the standard algorithm exhibits in
the static or extremely optically thin limit. Furthermore, the new solution is
not to replace, but to complement, the old solution.

The threshold at which we switch between solution strategies is somewhat
arbitrary; our choice is $\zeta=\num{e-5}$ as the standard algorithm has not
yet shown instability above it. If double-precision floating-point numbers are
used, the machine epsilon is $2^{-52}\approx\num{2.22e-16}$, so the iterative
step should be performed at least four times.

\section{IR initial condition}
\label{sec:radiative initial condition}

Because the time-dependent \ac{RT} module of Athena operates on $I_\su{IR}$
rather than $J_\su{IR}$ and $\vec H_\su{IR}$, we must convert $E^0_\su{IR}$
provided by the initial condition to $I_\su{IR}$. Inside the optically thick
torus body, the \ac{IR} specific intensity in the fluid frame can be found in
the \ac{FLD} approximation as \citep{1981ApJ...248..321L}
\begin{equation}
I^0_\su{IR}(\uvec n^0)\eqdef\frac c{4\pi}|E^0_\su{IR}|\mathcal R^{-1}|
  (\coth\mathcal R-\uvec m^0\cdot\uvec n^0)^{-1};
\end{equation}
it follows that
\begin{equation}
\vec H^0_\su{IR}=
  \frac c{4\pi}|E^0_\su{IR}|(\coth\mathcal R-\mathcal R^{-1})\,\uvec m^0.
\end{equation}
Here $\mathcal R\eqdef\norm{\grad
E^0_\su{IR}}/(\rho|\kappa_\su{IR}|E^0_\su{IR})$ is the Knudsen number for
radiation diffusion, and $\uvec m^0\eqdef-\grad E^0_\su{IR}/\norm{\grad
E^0_\su{IR}}$. Geometrically speaking, if we draw arrows $\uvec n^0$ from the
origin with lengths proportional to $I^0_\su{IR}(\uvec n^0)$, the envelope is a
prolate ellipsoid with ellipticity $\tanh\mathcal R$ and one focus at the
origin. We impose the additional constraint that $0\le\tanh\mathcal R\le0.95$;
the ceiling makes radiation less unidirectional in optically thin regions so
that at least a few grid rays carry finite specific intensity. The specific
intensity is then boosted to the observer frame by
\begin{equation}
I_\su{IR}(\uvec n)=I^0_\su{IR}(\uvec n^0)|
  \left[\frac{(1-v^2/c^2)^{1/2}}{1-\uvec n\cdot\vec v/c}\right]^4.
\end{equation}
We remarked after \cref{eq:radiative transfer} that $I_\su{IR}$ is a
frequency-integrated quantity, which explains why the exponent is four, not
three. Note that the \ac{FLD} approximation is used merely to define the
initial condition; it is not used to solve \cref{eq:radiative
transfer,eq:radiative zeroth moment,eq:radiative first moment}.

\section{Orbital velocity profile of a hydrostatic, radiation-supported torus}
\label{sec:velocity profile}

The force balance of a hydrostatic torus supported by \ac{IR} radiation against
the gravity of a point mass is expressed in
\begin{equation}
-\grad\left(-\frac{GM}r\right)+\frac{\kappa_\su{IR}}c|\vec F_\su{IR}+
  \frac{v_\phi^2}R\,\uvec e_R=\vec0.
\end{equation}
The equation is solved together with the constraint of \ac{IR} radiation energy
conservation, $\divg\vec F_\su{IR}=0$, and the assumption that $\kappa_\su{IR}$
is not a strong function of position. A similar equation has been solved by
\citet{2007ApJ...661...52K} under axisymmetry; here we present a more intuitive
approach. Because the gravitational and radiative terms are both
divergence-free, the same must also be true for $(\smash{v_\phi^2}/R)\,\uvec
e_R$. The only radial and divergence-free vector field is
$C(\phi,z)|R^{-1}\,\uvec e_R$ for some function $C(\phi,z)$, hence $v_\phi$ is
a constant over $R$. If we further restrict $v_\phi$ to be axisymmetric, we can
write
\begin{equation}
v_\phi(R,z)=j_\su{in}(z)|\left(\frac{GM}{R_\su{in}}\right)^{1/2}.
\end{equation}
Here $j_\su{in}(z)$ is some dimensionless function that measures the shortfall
of orbital velocity at $R=R_\su{in}$ from Keplerian as a consequence of
radiative support, so we have $0\le j_\su{in}\le 1$.

\end{appendices}

\ifaastex{\bibliography{torus}}{\printbibliography}

\end{document}